\documentclass[fleqn,usenatbib]{mnras}

\usepackage{graphicx}
\usepackage{xcolor}
\usepackage{comment}

\newcommand{\xigg}{\xi_{\rm{gg}}}
\newcommand{\dgg}{\Delta_{\rm{gg}}}
\newcommand{\xigl}{\xi_{\rm{g}\lambda}}
\newcommand{\xill}{\xi_{\lambda\lambda}}
\newcommand{\meanl}{\langle\lambda\rangle}
\newcommand{\lala}{\lambda\lambda}
\newcommand{\kk}{\kappa\kappa}
\newcommand{\lk}{\lambda\kappa}
\newcommand{\kl}{\kappa\lambda}
\newcommand{\insitu}{\emph{in situ} }

\newcommand{\mpch}{\,h^{-1}{\rm{Mpc}}}
\newcommand{\kpch}{\,h^{-1}{\rm{kpc}}}
\newcommand{\mpc}{\,{\rm{Mpc}}}

\newcommand{\msol}{{\,\rm{M}}_{\odot}}
\newcommand{\parsec}{{\rm{pc}}}
\newcommand{\dgr}{\delta_{{\rm{g}}, S}}
\newcommand{\reff}{R_{\rm{eff}}}
\newcommand{\rvir}{R_{\rm{vir}*}}

\usepackage{threeparttable}
\usepackage{tabulary}
\usepackage{tablefootnote}

\usepackage{newtxtext,newtxmath}
\usepackage[T1]{fontenc}

\DeclareRobustCommand{\VAN}[3]{#2}
\let\VANthebibliography\thebibliography
\def\thebibliography{\DeclareRobustCommand{\VAN}[3]{##3}\VANthebibliography}

\usepackage{graphicx}	% Including figure files
\usepackage{amsmath}	% Advanced maths commands
\title[Intrinsic Size Correlations]{Intrinsic correlations of galaxy sizes in a hydrodynamical cosmological simulation}

\author[H. Johnston et al.]{Harry Johnston,$^{1}$\thanks{E-mail: h.s.johnston@uu.nl}
Dana Sophia Westbeek,$^{1}$
Sjoerd Weide,$^{1}$
Nora Elisa Chisari,$^{1}$
Yohan Dubois,$^{2}$
\newauthor
Julien Devriendt,$^{3}$
Christophe Pichon$^{2,4}$
\\
$^{1}$Institute for Theoretical Physics, Utrecht University, Princetonplein, 3584 CC, Utrecht, The Netherlands\\
$^{2}$CNRS and Sorbonne Universit\'e, UMR 7095, Institut d'Astrophysique de Paris, 98 bis, Boulevard Arago, F-75014 Paris, France.\\
$^{3}$Sub-department of Astrophysics, University of Oxford, Keble Road, Oxford OX1 3RH, UK.\\
$^{4}$IPhT, DRF-INP, UMR 3680, CEA, L'Orme des Merisiers, Bât 774, 91191 Gif-sur-Yvette, France.
}

\date{Accepted 2023 January 17. Received 2022 January 17; in original form 2022 September 22.}

\pubyear{2023}

\begin{document}
\label{firstpage}
\pagerange{\pageref{firstpage}--\pageref{lastpage}}
\maketitle

\begin{abstract}
Residuals between measured galactic radii and those predicted by the Fundamental Plane (FP) are possible tracers of weak lensing magnification. However, observations have shown these to be systematically correlated with the large-scale structure.
We use the Horizon-AGN hydrodynamical cosmological simulation to analyse these intrinsic size correlations (ISCs) for both elliptical (early-type) and spiral (late-type) galaxies at $z=0.06$. 
We fit separate FPs to each sample, finding similarly distributed radius residuals, $\lambda$, in each case. We find persistent $\lambda\lambda$ correlations over three-dimensional separations $0.5-17\,h^{-1}{\rm{Mpc}}$ in the case of spiral galaxies, at $>3\sigma$ significance.
When relaxing a mass-selection, applied for better agreement with galaxy clustering constraints, the spiral $\lambda\lambda$ detection strengthens to $9\sigma$; we detect a $5\sigma$ density-$\lambda$ correlation; and we observe intrinsically-large spirals to cluster more strongly than small spirals over scales $\lesssim10\,h^{-1}{\rm{Mpc}}$, at $>5\sigma$ significance. Conversely, and in agreement with the literature, we observe lower-mass, intrinsically-small ellipticals to cluster more strongly than their large counterparts over scales $0.5-17\,h^{-1}{\rm{Mpc}}$, at $>5\sigma$ significance.
We model $\lambda\lambda$ correlations using a phenomenological non-linear size model, and predict the level of contamination for cosmic convergence analyses. We find the systematic contribution to be of similar order to, or dominant over the cosmological signal. We make a mock measurement of an intrinsic, systematic contribution to the projected surface mass density $\Sigma(r)$ and find statistically significant, low-amplitude, positive (negative) contributions from lower-mass spirals (ellipticals), which may be of concern for large-scale ($\gtrsim\,7\,h^{-1}$ Mpc) measurements.
\end{abstract}


%--------------------------------------------------------------------

\section{Introduction}
Weak gravitational lensing is typically studied through estimation of the complex gravitational shear field $\gamma$, traced by the measurable ellipticities of source galaxies over a range of redshifts. Cosmic shear correlators defined with this evolving field, and the evolving galaxy density field, yield information describing the amount and clustering of matter in the late-time Universe \citep{Hikage2018,Asgari2020b,Heymans2020,Secco2022,Abbott2022}.

The other component of weak lensing is the magnification sourced by the scalar convergence field $\kappa$, often probed via lensing of the cosmic microwave background \citep[CMB;][]{PlanckCollaboration2018a,Fang2021}.  In photometric galaxy weak lensing surveys, the projected surface mass density $\Sigma \propto \kappa$ at a given redshift magnifies sources behind it. Effectively, it changes the observed flux coming from a distant source and the solid angle it subtends, while preserving its surface brightness. The increase in flux promotes galaxies across the survey detection threshold. Simultaneously, the background number density is diluted by the increase in solid angle. The relative strength of these competing effects is determined by the faint-end slope of the survey luminosity function \citep{Bartelmann1999}; how many faint objects are `waiting' to be promoted across the flux limit.

Magnification can induce galaxy number density fluctuations and correlations that must be correctly modelled in order to avoid catastrophic biases in cosmological parameter inference \citep{Cardona2016,Hoekstra2017,Thiele2020,Unruh2020,Duncan2021,Mahony2021}. However, it also has potential as an aide to weak lensing halo mass calibration \citep{Rozo2010,Hildebrandt11,Umetsu11,Hildebrandt2013,Ford14,Duncan16}, and as a cosmological probe; various techniques have been developed to detect magnification correlations via measured galaxy sizes, magnitudes, redshifts, number densities, and even shears \citep{Myers2005,Scranton2005,Hildebrandt2009,Menard2010,Morrison2012,Schmidt2012,Alsing2015,Garcia-Fernandez2018,Liu2021}.

Whilst recent work suggests that gains in the precision of cosmological parameter inference from the inclusion of magnification in cosmic shear analyses are modest when galaxy clustering is also included \citep{Duncan2014,Lorenz2018,Duncan2021,Mahony2021}, the potential for magnification analyses independent from cosmic shear systematics offers a valuable consistency test \citep{Hildebrandt2009,VanWaerbeke2010,Alsing2015,Ghosh2020}, and future space surveys could enable magnification to approach the statistical power of cosmic shear \citep{Casaponsa13,Heavens2013}.

Some works have explored the possibility of measuring weak lensing magnification by correlating the sizes of galaxies as characterised by residuals with respect to the Fundamental Plane \citep{Bertin2006,Huff2011,Freudenburg2020}. The Fundamental Plane \citep{Djorgovski1987} is a tight set of scaling relations, originating from the virial theorem. Assuming homologous mass-to-light ratios and constant surface brightness, one derives the Faber-Jackson relation between luminosity and velocity dispersion, and the Fundamental Plane (FP), relating the radius, velocity dispersion, and surface brightness of elliptical (early-type) galaxies. Breakdowns of these assumptions result in the well-studied `tilting' of the FP \citep{Djorgovski1987,Bernardi2003,Hyde2009,Saglia2010,Cappellari2013a,Cappellari2013,Saulder2013}.

\cite{LaBarbera2010} had previously shown that the best-fitting FP for a given sample of elliptical galaxies is sensitive to the local density contrast, and \cite{joachimi} and \cite{Singh2020} made detections of auto- and cross-correlations between FP radius residuals and the large-scale structure for elliptical galaxies from SDSS DR8 \citep{Saulder2013}, and BOSS CMASS \& LOWZ \citep{Alam2015}, respectively.

Intrinsic galaxy alignments -- wherein tidal forces orient galaxies toward local density peaks in three dimensions -- are thought to induce an orientation-dependent scatter in the galaxy size distribution, after shapes are projected onto the two-dimensional surface of the celestial sphere \citep{Hirata2009,Martens2018}. However, \cite{Singh2020a} showed that these were not sufficient to explain observed density-FP residual correlations, which must therefore feature some other physical or systematic contributions.

In analogy to the intrinsic alignments \citep[IA;][]{Catelan2001,Hirata2004a} of galaxies as a contaminant to cosmic shear, intrinsic spatial correlations between galaxy sizes, or between sizes and the density field, could be mistakenly attributed to lensing magnification, and thus bias the cosmological interpretations of measured observables \citep{Ciarlariello2015,Ciarlariello2016}.

Moreover, intrinsic size correlations (ISCs) could further mimic magnification in promoting galaxies across detection thresholds. A hybrid of the lensing-induced size bias \citep{Schmidt2009}, and the lensing-independent tidal alignment bias \citep{Hirata2009,Martens2018}, an \emph{intrinsic size bias} would be induced by aperture selections -- a density-dependent selection effect with the potential to contaminate all measurable galaxy statistics \citep{Schmidt2009a}. To our knowledge, such an effect has not been explicitly studied in the literature, though developments in effective field theories offer possible avenues to do so \citep[see e.g.][]{Agarwal2021}.

Besides the possibility of intrinsic contamination of lensing statistics derived from galaxy sizes, intrinsic galaxy size correlations are themselves of interest for astrophysics and cosmology. The scaling of galaxy sizes with their environments and other properties, as it relates to the divergent dynamics of early- and late-type objects, and their interplay with galaxy merger events, are all promising laboratories for studies of galaxy formation and evolution \citep{Shen2003,Governato2007,Vale2008,Naab2010,Oser2010,Newman2012,Dubois2013,Dubois2016,Kravtsov2013,Cappellari2013,Welker2014,Welker2017}.

This work uses data from the hydrodynamical Horizon-AGN simulation \citep{Horizon} to explore the landscape of intrinsic size correlations with the benefit of precise determination of galaxy properties, and in the absence of gravitational lensing. We characterise intrinsic size distributions according to deviations from the fitted Fundamental Plane, similarly to \cite{joachimi} and \cite{Singh2020}, and we extend this concept to include an FP for spiral (late-type) galaxies, inspired by \cite{Shen2002} who investigated spiral FPs motivated by the Tully-Fisher relation. This is a significant extension, as one can assume that spiral galaxies will dominate the deep samples utilised for studies of lensing magnification \citep{Huff2011}.

We describe our simulated data in Sect. \ref{sec:simulation}. Our Fundamental Planes and definitions of intrinsic sizes are detailed in Sect. \ref{sec:fundamental_planes}. In Sect. \ref{sec:intrinsic_size_correlations} we outline our estimators for intrinsic size and galaxy density correlations in the simulation box, and our methods for predicting intrinsic contamination of magnification signals. Sect. \ref{sec:discussion} discusses our measured correlations, contamination predictions, and their implications, and our concluding remarks are presented in Sect. \ref{sec:conclusions}.

Throughout, we work with the $\Lambda$CDM cosmology of the Horizon-AGN simulation: $\Omega_{\rm{m}}=0.272,\,\Omega_{\rm{b}}=0.045,\,\Omega_{\Lambda}=0.728,\,\sigma_8=0.81,\,H_0=70.4\,{\rm{kms^{-1}Mpc^{-1}}},\,n_{\rm{s}}=0.967$. All quoted distances are comoving distances, though applications of a factor $h$ (the dimensionless Hubble parameter) can differ, and are specified by the quoted units.

%--------------------------------------------------------------------
   
\section{Simulation}
\label{sec:simulation}

\begin{figure*}
    \centering
    \includegraphics[width=\textwidth]{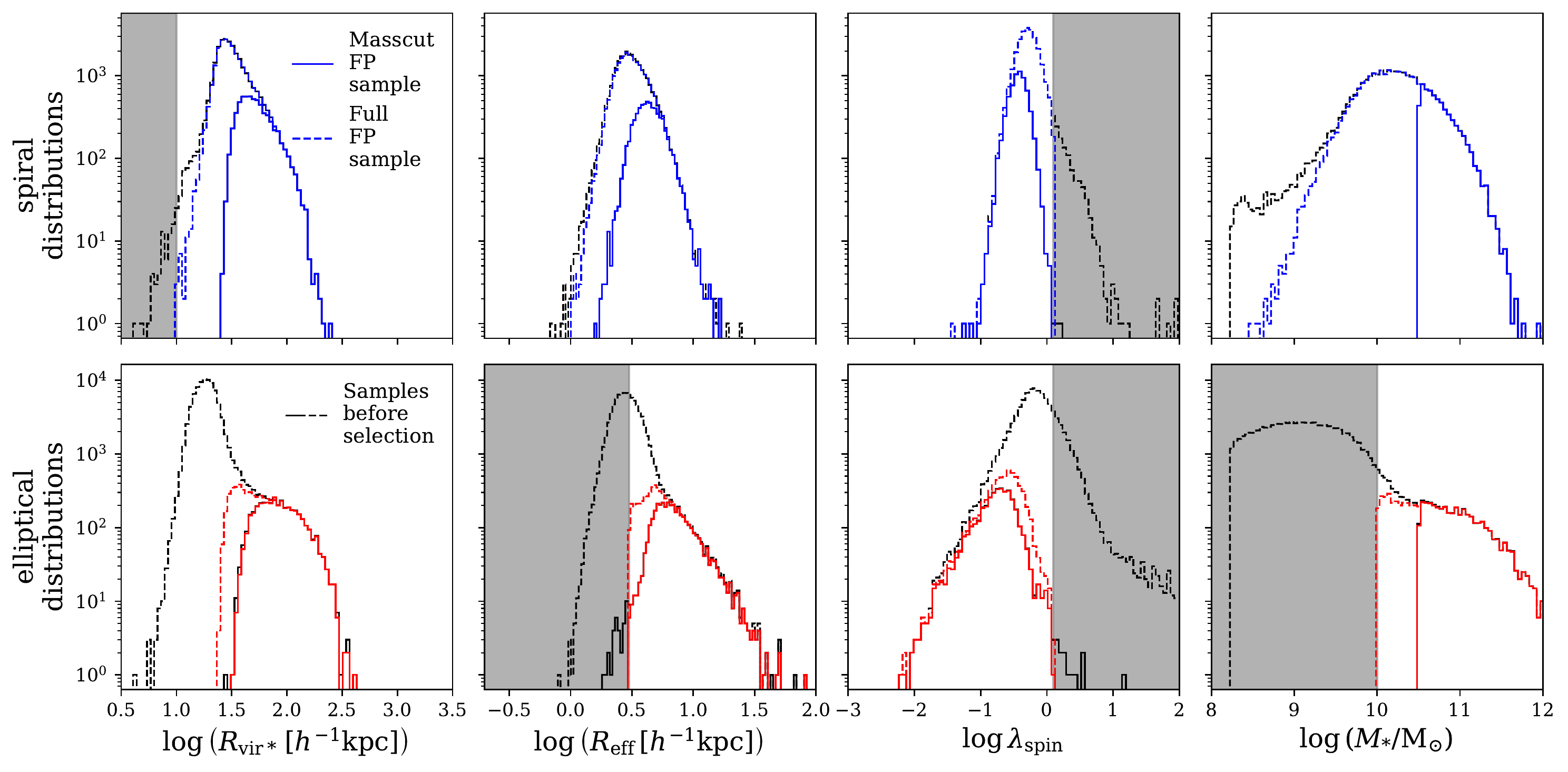}
    \caption{Distributions of some spiral (\emph{top}) and elliptical (\emph{bottom}) galaxy properties in the Horizon-AGN $z=0.06$ snapshot, selected according to a $V/\sigma=0.6$ boundary. Solid-line (dashed-line) histograms denote Masscut (Full) samples to be fitted with Fundamental Planes, after application of all selections (shown here as grey shading). Black histograms display the Masscut/Full galaxy populations before any additional selections (see Sect. \ref{sec:simulation}) on virial radii  (\emph{left}), effective radii (\emph{middle-left}), spin magnitudes (\emph{middle-right}), and stellar masses (\emph{right}). The mass-selection works to remove small-radius spirals and ellipticals from the samples, affecting $75\%$ and $40\%$ of selected objects, respectively.
    }
    \label{fig:distributions}
\end{figure*}

Horizon-AGN is a cosmological hydrodynamical simulation, run with a standard flat $\Lambda$CDM cosmology \{$\Omega_{\rm{m}}=0.272,\,\Omega_{\rm{b}}=0.045,\,\Omega_{\Lambda}=0.728,\,\sigma_8=0.81,\,H_0=70.4\,{\rm{kms^{-1}Mpc^{-1}}},\,n_{\rm{s}}=0.967$\} compatible with the constraints from WMAP7 \citep{Komatsu2011}. The box has a width of $L_{\rm{box}}=100\mpch$ and $1024^3$ dark matter particles, each having a mass of $8\times10^7\msol$. Run with the adaptive mesh refinement code {\sc{ramses}} \citep{Teyssier2002}, and an initial gas mass resolution of $10^7\msol$, the mesh is adaptively refined down to $\Delta{x}=1$ proper kpc. For more details on the simulation of gas cooling, star formation, and stellar and black hole feedback processes, see \cite{Horizon} and \cite{Dubois2016}.

At each redshift snapshot, galaxies are identified in the simulation by running the {\sc{AdaptaHOP}} subclump finder \citep{Aubert2004} against the stellar particle distribution. Galactic stellar masses are computed as the sum over all ($\geq50$) stellar particles attributed to a given galaxy by {\sc{AdaptaHOP}}. Galaxy luminosities are computed using single stellar population models \citep{Bruzual2003} with a Salpeter initial mass function. The flux contributed by each star depends on its mass, metallicity and age. Total fluxes are convolved with Sloan Digital Sky Survey filters \citep{Gunn06} without dust extinction to obtain absolute AB magnitudes and rest-frame galaxy colours (e.g., $g-r$).

Horizon-AGN has been shown to reproduce a host of observed stellar and galactic observations: luminosity and stellar mass functions; the main sequence of star formation; the rest-frame colour distribution from UV to infrared; and the cosmic star formation history \citep{Kaviraj}. \cite{Dubois2016} showed the simulation to agree with observations of galaxy size-stellar mass relations from 3D-HST and CANDELS \citep{VanDerWel2014}, and cited feedback from active galactic nuclei as important in achieving extended galactic profiles over cosmic evolution. \cite{Hatfield2019} measured galaxy clustering in Horizon-AGN lightcones designed to mimic VIDEO photometric observations \citep{Jarvis2013}, finding agreement for galaxies with stellar masses $M_{*}\gtrsim10^{10.5}\msol$.

The successful reproduction of trends in galaxy size and clustering is crucial to our work here, as we seek to take advantage of precisely-known, lensing-free simulated quantities to gather meaningful predictions for intrinsic size correlations in the real Universe. As such, we consider galaxy samples both with and without a cut to the stellar mass $M_{*}>10^{10.5}\msol$, and advise that the lower-mass results be taken with circumspection. Throughout this work, sample two-point correlations will feature the mass selection on (i) both, or (ii) neither of the samples. 

For all galaxies, the object's semi-major $a$ and semi-minor $b$ axes are computed in projection along the $z$-axis of the box \citep{Chisari2015}, and two-dimensional, anisotropic surface areas are defined as $\pi ab$. 

The velocity dispersion per galaxy $\sigma=\frac{1}{3}\sqrt{\sigma^{2}_{r} + \sigma^{2}_{\theta} + \sigma^{2}_{z}}$ is computed via the radial $\sigma_r$, tangential $\sigma_\theta$, and vertical $\sigma_z$ dispersion components, defined with respect to the angular momentum vector of each galaxy, and the rotational velocity $V=\bar{v}_{\theta}$ is the average of stellar tangential velocity components $v_{\theta}$ \citep{Dubois2016}. A `circular' velocity $V_{\rm{c}}$ is also defined, as the rotational velocity at the virial radius $\rvir$ -- the radius within which the virial theorem is satisfied by the stellar particles assigned to the object.

Spiral galaxies are rotationally supported, having coherent stellar motions and low stellar velocity dispersion, whilst pressure-supported ellipticals have high stellar velocity dispersion due to the random motions of stars. We thus follow previous works in defining the boundary between elliptical and spiral galaxies according to the ratio of rotational velocity $V$ to velocity dispersion $\sigma$. We divide the sample at $V/\sigma=0.6$ -- though this choice is fairly arbitrary, it is roughly where \cite{Horizon} saw the `spin-flip' occurring in Horizon-AGN, whereby the alignments of galaxies' spin axes with nearby filamentary orientations transitions from (spiral/low-mass) parallel to (elliptical/high-mass) perpendicular \citep{Welker2014,Codis2018,Bate2019,GaneshaiahVeena2019,Lee2021,Kraljic2021}. As a marker of a morphological transition brought about by mergers, and having dramatic implications for correlations between galaxy spins/shapes and local tidal fields, this seems a sensible place to separate our putative intrinsic size correlations into spiral and elliptical contributions. We note, however, that a more detailed morphological classification would be of interest for studies seeking to bridge the gap between the elliptical and spiral FPs, and for studies of ISCs and IA in general.

A unified Fundamental Plane for elliptical and spiral galactic radii may prove elusive, given the divergent dynamics of such systems. However, \cite{Ferrero2021} show with hydrodynamical simulations and observations that the Tully-Fisher and Faber-Jackson scaling relations can be unified upon consideration of the ratio of stellar-to-dark halo mass enclosed within an effective (i.e. stellar half-mass) radius, going on to demonstrate a unified stellar mass plane. This, along with constraints upon galaxy stellar mass-size scaling relations \citep{Kawinwanichakij2021,Nedkova2021,Rodriguez2021}, and the identification of other tight scaling relations for spiral galaxies \citep{Lagos2016,Matthee2018,ManceraPina2021}, should place a generalised FP for the radii of galaxies, agnostic of morphological type, within the realm of possibility.

If a unified plane were to be identified, besides offering further insights into galaxy formation and evolution via FPs and ISCs, one might achieve a reduction in the number of parameters required to model ISCs for magnification studies, and open up the possibility for size-based convergence analyses utilising both morphological types, with colour-split FPs then offering a cross-checking mechanism. In this work, for simplicity, we define separate FPs for ellipticals and for spirals.

\subsection{Elliptical Fundamental Plane properties}
\label{sec:elliptical_properties}

The effective radius $\reff$ in this work is equal to the geometric mean of three half-mass radii, each computed after projection of an object's stellar particle distribution along a Cartesian axis of the simulation box \citep{Dubois2016}. \cite{Rosito2020} use a slightly different definition in their study of FPs in Horizon-AGN, where elliptical radii are taken as the three-dimensional radii containing half of each object's stellar mass -- these estimates are nonetheless comparable, inhabiting a similar dynamic range.

The surface brightness $I$ is computed as the object luminosity (the sum of stellar particle luminosities) per unit area  ($A=\pi a b$). We note that our combination of projected and three-dimensional quantities is likely to introduce some additional scatter and discrepancies with respect to the FPs of \cite{Rosito2020}, who avoided projections. They also explored FPs replacing the surface brightness $I$ with the surface mass density $\Sigma$ -- we shall refer to these planes as L-FP and $\Sigma$-FP, respectively, in forthcoming sections.

The final parameter of the elliptical FP is the stellar velocity dispersion $\sigma$ defined above.

\subsection{Spiral Fundamental Plane properties}
\label{sec:spiral_properties}

\cite{Shen2002} worked from the Tully-Fisher (TF) relation, which describes the positive scaling of spiral galaxies' luminosity $L$ with the circular velocity $V_{\rm{c}}$. Monte Carlo sampling their detailed disc dynamics model in search of a third variable for the FP, they found that the shape of the rotation curve, and the disc scale-length $R_{\rm{d}}$, were correlated with the scatter around the TF relation. We neglect to make specific estimations of $R_{\rm{d}}$ for our simulated galaxies, finding the virial radius of stellar particles $\rvir$ to be similarly correlated with the TF scatter. We thus make use of $\rvir$ for our spiral FP, as a simple addition to the TF relation, and advise that follow-up work use estimates of more observationally tractable quantities such as the disc-scale length.
 
We note that the two radii under consideration have different dynamic ranges, with a typical empirical ratio $\reff\sim0.1\rvir$. Both radii are measured using only the stellar particles assigned to an object by {\sc{AdaptaHOP}}, but the effective radius is an averaged half-mass radius (Sect. \ref{sec:elliptical_properties}), whilst the virial radius is defined such that the motions of enclosed particles satisfy the virial theorem. We thus emphasise that the intrinsic sizes we are to define (Sect. \ref{sec:intrinsic_sizes}) are probing inner-galactic radii for elliptical galaxies, $\sim3-50\kpch$, and outer-galactic radii, $\sim10-300\kpch$, for spirals.

Using abundance matching techniques and observational galaxy data, \cite{Kravtsov2013} found galaxy half-mass and total (i.e. stellar plus dark matter) virial radii to scale almost linearly, with a power-law slope of $\sim0.95$, across two decades of radius and eight decades of stellar mass -- thus including all morphological types -- and with a scatter of $\sim0.2$ dex. 

Our Horizon-AGN samples display similar scatters in the $\reff-\rvir$ relation, but shallower power-law slopes of $\sim0.5-0.6$ for spiral samples, as well as on-average larger $\reff$ and expectedly smaller $\rvir$ (given that our virial radii are estimated only from the stellar particle distribution), resulting in a normalisation $\sim10\times$ larger than seen by \cite{Kravtsov2013}. Concurrently, we find that $\rvir$ enables reasonable fits of a spiral Fundamental Plane (Sect. \ref{sec:fundamental_planes}). This complicates the interpretation of our measured intrinsic size correlations, which we shall attempt to make clear in the coming sections.

The remaining parameters for the spiral FP are the circular velocity $V_{\rm{c}}$ defined above, and the absolute $r$-band magnitude $M_r$.

\subsection{Additional selections}
\label{sec:additional_selections}

In refining our FPs for spiral and elliptical galaxies, some objects were found to be poorly described. These included:

\begin{itemize}
    \item Elliptical galaxies with effective radii $\reff<3\kpch$, approaching the resolution limit of the simulation;
    \item Elliptical galaxies with stellar masses $M_{*}<10^{10} \msol$;
    \item Spiral galaxies with stellar virial radii $\rvir<10\kpch$;
    \item Galaxies with very large spin magnitudes \citep[see][Eq. 2]{Aubert2004}, which were found on inspection to be recently merged/merging structures that contaminate lower-mass samples.
\end{itemize}
We exclude these objects from our FP samples, as each contributes to heavily increased scatter, tilting, curving, or other irregular structures on the fitted planes. We also cut away the aforementioned lower-mass galaxies with $M_{*}<10^{10.5}\msol$ \citep{Hatfield2019} prior to fitting a more conservative set of FPs (see Sec. \ref{sec:simulation}), which we denote as `Masscut FP', as opposed to the `Full FP' samples which include those lower-mass objects.

Fig. \ref{fig:distributions} shows the resulting spiral (top) and elliptical (bottom) galaxy property distributions, with solid-line histograms giving the Masscut samples, and dashed-line histograms the Full samples. Colours denote the galaxies that we fit with FPs after all selections (shown as grey shading) are applied, whilst black-lined histograms show the total Masscut/Full sample populations before any additional selections. One sees that the confluence of stellar mass, radius, and spin selections serves to exclude populations of small, low-mass, fast-spinning galaxies, which are known to be over-produced in the simulation \citep{Kaviraj2017}. 

\begin{figure}
    \centering
    \includegraphics[width=0.47\textwidth]{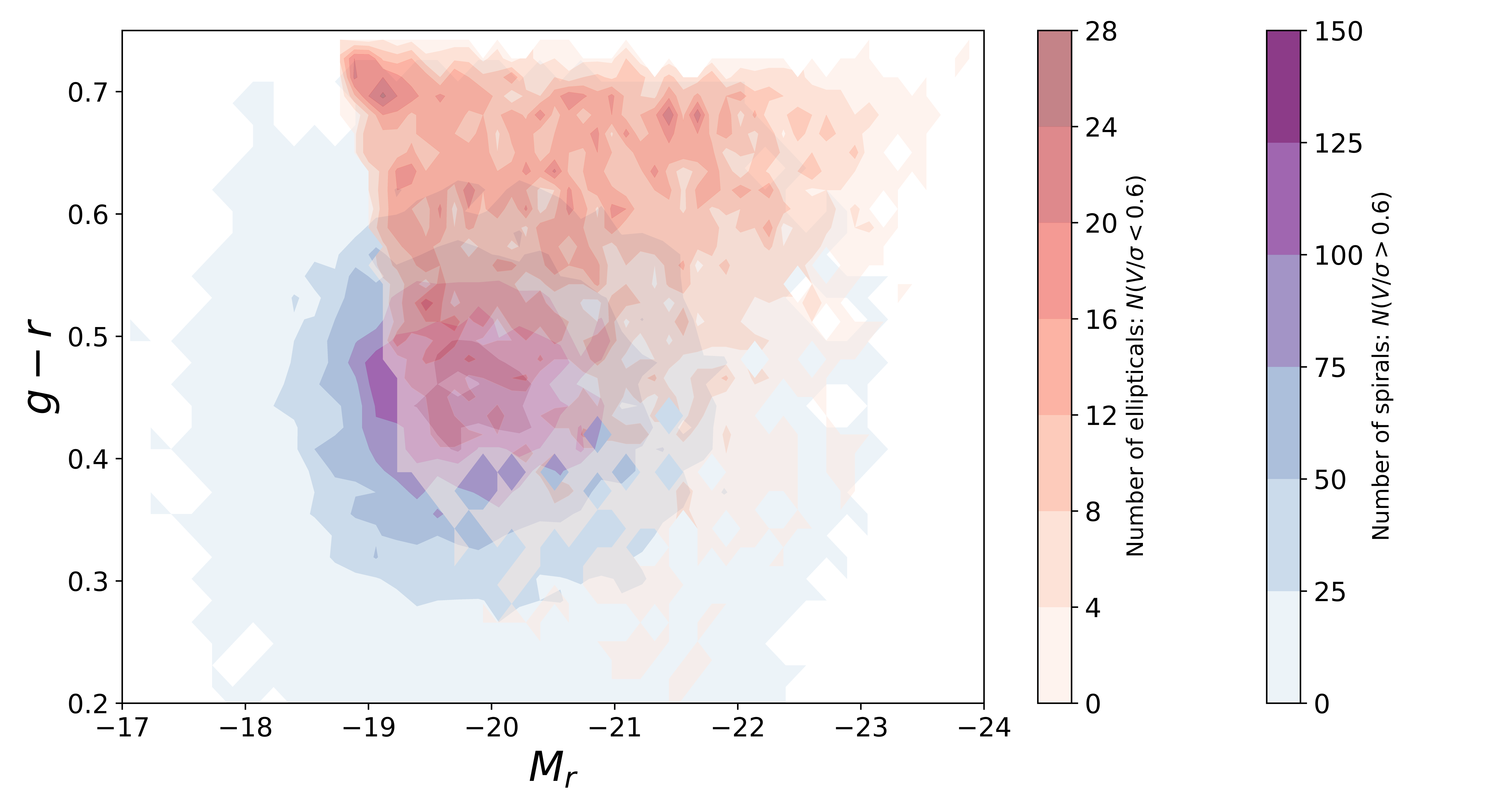}
    \caption{Colour-magnitude diagrams for the full sample of spirals (blue) and ellipticals (red) in the Horizon-AGN $z=0.06$ snapshot, selected according to a $V/\sigma=0.6$ boundary. The colour-magnitude diagram clearly indicates the presence of a ``blue cloud'' and a ``red sequence''. The $V/\sigma$ cut serves to separate the populations efficiently, though some of the ellipticals still have significantly blue colours.
    }
    \label{fig:cmag}
\end{figure}

Fig. \ref{fig:cmag} shows the colour-magnitude diagram for the two full samples. Spirals occupy the ``blue cloud'' and ellipticals have mostly redder colours, clearly defining a ``red sequence''. There is a significant elliptical fraction that overlaps with the blue cloud given our selection cuts. This is a known issue in cosmological hydrodynamical simulations. Despite colours largely matching observations, as demonstrated in \cite{Kaviraj2017}, it is common to find bluer colours than in observations. Nevertheless, it is clear that despite some overlap, the $V/\sigma$ cut works to separate the two populations efficiently. This is also evidenced in terms of the quality of the fits of the fundamental relations of each sample, which we describe in Section \ref{sec:fundamental_planes}.

%--------------------------------------------------------------------

\section{Fundamental Planes}
\label{sec:fundamental_planes}

\cite{Rosito2020} studied the morphology, kinematics, and scaling relations of elliptical (early-type) galaxies in Horizon-AGN. They found tight FPs with no clear redshift evolution for $z\leq3$, and that energetic feedback from active galactic nuclei (omitted from the initially identical companion simulation, Horizon-noAGN) is necessary for the reproduction of several observations, including the Fundamental Plane. Our elliptical FPs are not directly comparable to those of \cite{Rosito2020}, who estimate a three-dimensional half-mass radius (different to our $\reff$, defined by \citealt{Dubois2016}) and use this to compute circularised surface areas. However, we do see a qualitatively similar plane, with equivalent signs in the coefficients, a comparable root-mean-square (rms) of residuals, and some minor asymmetry about the 1:1 relation (see Sec. \ref{sec:fundamental_planes_results}). Given that \cite{Rosito2020} observed no redshift evolution of the FP, we limit our analysis here to a single simulation snapshot at $z=0.06$, though we note that follow-up work incorporating higher-$z$ snapshots would be desirable.

We define our FPs similarly to \cite{joachimi} and \cite{Rosito2020}, but omit the redshift dependence employed by the former, since we consider only a single simulated redshift snapshot. We retain the surface brightness $I$ for our FP, in contrast to \cite{Rosito2020}, who showed that the `L-FP' (surface brightness FP) for their Horizon-AGN sample had a less symmetric scatter about the $1:1$ relation, as compared with an FP using the surface stellar mass density $\Sigma$ ($\Sigma$-FP).

We find the opposite to be true for our samples, most likely as a consequence of the discrepant projected/three-dimensional quantities already discussed, though we also have minor differences in selections (e.g. \citealt{Rosito2020} selected central galaxies according to numbers of stellar particles, whilst we make various cuts against stellar mass). Whilst the L-FP/$\Sigma$-FP difference is small at the level of the planes (hence we neglect to investigate in great detail), we shall see in Sect. \ref{sec:intrinsic_size_correlations_results} that the $\Sigma$-FP erases intrinsic size correlations as seen by the L-FP.

For elliptical galaxies, the FP is then given by
\begin{equation}
    \log{}\,\reff = a\,\log\sigma + b\,\log{}I + c,
    \label{eq:FP_elliptical}
\end{equation}
where $\reff$ is the object's effective radius, $\sigma$ is the stellar velocity dispersion, and $I$ is the surface brightness (for which \citealt{Rosito2020} substituted surface stellar mass density $\Sigma$). Throughout this work, we shall denote the base-10 logarithm as `$\log$', and the natural logarithm as `$\ln$'.

Following \cite{Shen2002}, Eq. 13, we use a radius estimate to tighten the TF relation between luminosity $L$ (or absolute magnitude) and circular velocity $V_{\rm{c}}$. Our spiral galaxy FP is thus given as
\begin{equation}
    \log{}\,\rvir = \alpha\,M_r + \beta\,\log{}V_{\rm{c}} + \gamma \,,
    \label{eq:FP_spiral}
\end{equation}
where $M_r$ is the simulated $r$-band absolute magnitude, and the circular velocity $V_{\rm{c}}$ is estimated as the rotational velocity at the virial radius $\rvir$, which we use in lieu of the disc scale length $R_{\rm{d}}$ employed by \cite{Shen2002} (see Sect. \ref{sec:spiral_properties}).

We normalise each FP parameter to its median value before fitting the coefficients $a,\,b,\,c,\,\alpha,\,\beta,\,\gamma$ to each defined galaxy sample via ordinary linear regression\footnote{See e.g. \cite{Magoulas2012,Said2020,Howlett2022} for more complex, censored 3D Gaussian models for FP fitting, which are useful in the context of measurement errors. For our simulated, effectively noiseless quantities, we assume that linear regression will suffice.}, and discuss the resulting FPs in Sec. \ref{sec:fundamental_planes_results}.

\subsection{Intrinsic sizes}
\label{sec:intrinsic_sizes}

Following \cite{joachimi}, we characterise the `intrinsic sizes' of galaxies according to residuals between their measured radii and corresponding predictions from fitted Fundamental Planes. The dimensionless intrinsic size parameter $\lambda$ is given as
\begin{equation}
 \lambda\equiv \ln\left(\frac{R_{\rm I}}{R_{\rm FP}}\right)\approx\frac{R_{\rm{I}}}{R_{\rm{FP}}}-1\,,
 \label{eq:deflambda}
\end{equation}
for the measured intrinsic radius $R_{\rm{I}}$, which we take as $\reff$ or $\rvir$, for ellipticals and spirals, respectively, and the predicted radius $R_{\rm{FP}}$ from the fitted FP. We will label the set of galaxies whose radii are predicted to be larger than they are in the simulation as $\lambda_+$, and those which are smaller as $\lambda_-$.

In this work, we are primarily interested in the intrinsic variability of galaxy sizes as a possible contaminant to the size fluctuations that one might attribute to the weak lensing convergence field $\kappa$. If we consider that lensing magnification operates on the intrinsic radius $R_{\rm{I}}$ to produce the observed radius of a galaxy $R_{\rm{O}}$, then we can explicitly define the latter as
\begin{equation}
    R_{\rm{O}} = R_{\rm{I}} (1+\kappa) \approx R_{\rm{FP}} (1+\lambda) (1+\kappa)\,,
    \label{sec:observed_radius_expanded}
\end{equation}
thus motivating our search for spatially-correlated, \emph{lensing-independent} size fluctuations $\lambda$ that could bias size-based estimates of the convergence $\kappa$ (see Sect. \ref{sec:intrinsic_size_correlations}).

\subsection{Fundamental Plane results}
\label{sec:fundamental_planes_results}

\begin{table*}
    \centering
    \begin{tabular}{lcccccc}
\hline
\hline
FP/lens sample & $N$ & $\langle L \rangle/L_{\rm{piv}}$ & $a\,|\,\alpha$ & $b\,|\,\beta$ & $c\,|\,\gamma$ & $\sigma_{\rm{FP}}$ \\
\hline
Elliptical FP & 6254 & 0.55 & $1.210\pm0.093$ & $-0.340\pm0.055$ & $0.729\pm0.013$ & $0.0737$ \\
\hline
Elliptical FP ($>10^{10.5}\,\rm M_{\odot}$) & 3684 & 0.84 & $1.041\pm0.142$ & $-0.380\pm0.072$ & $0.837\pm0.017$ & $0.0708$ \\
\hline
Spiral FP & 26215 & 0.24 & $-0.217\pm0.016$ & $-0.587\pm0.105$ & $-1.488\pm0.006$ & $0.0866$ \\
\hline
Spiral FP ($>10^{10.5}\, \rm M_{\odot}$) & 6394 & 0.59 & $-0.286\pm0.029$ & $-1.289\pm0.217$ & $-1.295\pm0.013$ & $0.0877$ \\
\hline
Lens & 7479 & 0.49 & $--$ & $--$ & $--$ & $--$ \\
\hline
Lens ($>10^{10.5}\,\rm M_{\odot}$) & 3741 & 0.84 & $--$ & $--$ & $--$ & $--$ \\
\hline
\hline
    \end{tabular}
    \caption{Sample details and Fundamental Plane constraints (if applicable) for elliptical, spiral (size tracers), and lens (density tracer) galaxy samples defined in the Horizon-AGN simulation. Columns give the sample, the number of galaxies, the ratio of the mean sample luminosity to a pivot luminosity corresponding to an absolute magnitude $M=-22$, the FP coefficients for elliptical (Eq. \ref{eq:FP_elliptical}) and spiral (Eq. \ref{eq:FP_spiral}) FPs, and the root-mean-square deviation from the FP, respectively.
    }
    \label{tab:sample_details}
\end{table*}

\begin{figure*}[h]
    \centering
    \includegraphics[width=\textwidth]{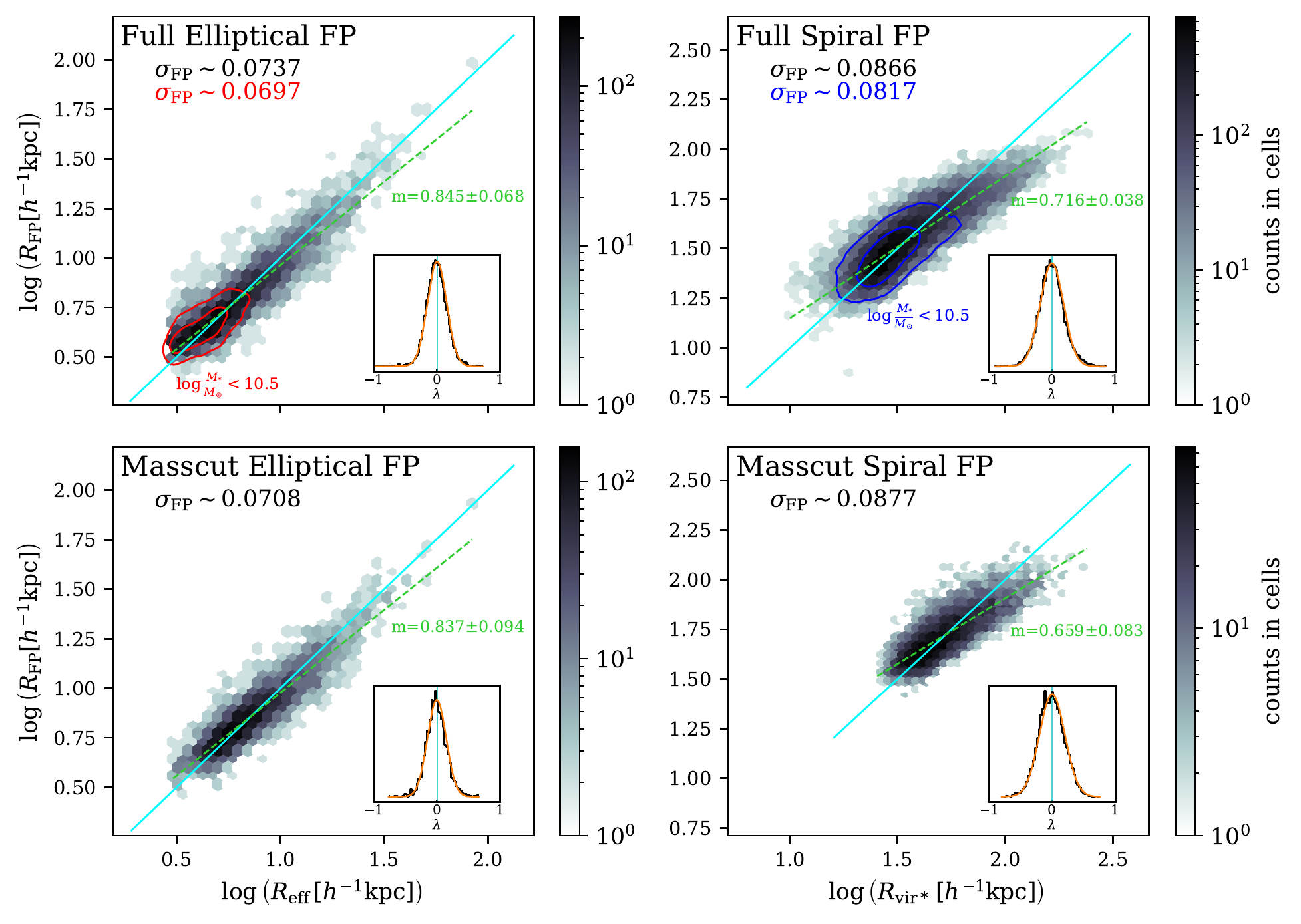}
    \caption{Fundamental Planes fitted to elliptical (\emph{left}) and spiral (\emph{right}) galaxies selected from the Horizon-AGN simulation (Sect. \ref{sec:simulation}). Hexagonally-binned two-dimensional histograms show the (log) $R_{\rm{FP}}-R_{\rm{eff/vir}}$ distributions for the Full Fundamental Planes (\emph{top}), containing all galaxies that passed our selections, with the coloured contours marking $68\%$ and $95\%$ of the lower-mass galaxies that are lost from the Masscut Fundamental Plane samples (\emph{bottom}). The $1:1$ relations are shown as solid cyan lines, and can be compared with linear fits to all points on each plane, shown as dashed green lines, and accompanied by the fitted gradients. Inset figures show the Fundamental Plane radius residuals $\lambda\equiv\ln{\frac{R_{\rm{I}}}{R_{\rm{FP}}}}$, overlain with Gaussian distributions (orange). The root-mean-square deviations $\sigma_{\rm{FP}}={\rm{rms}(\lambda)/\ln10}$ are given in black for the respective plane, and in colour for the subsets of lower-mass objects on the Full Fundamental Planes. Each of the fitted planes exhibits some degree of tilting, with spiral and Masscut planes most affected.
    }
    \label{fig:fittedFPs}
\end{figure*}

Our fitted FPs (shown in Fig. \ref{fig:fittedFPs}) yield residuals $\lambda$ (inset axes) that are closely comparable with Gaussian distributions (orange curves), with means $\meanl$ of $\mathcal{O}\left[10^{-10}\right]$ or smaller. Viewing the planes edge-on, we observe tilting in each, such that (small) large measured radii are (over-) under-predicted by the FPs; we characterise the degree of tilting by fitting a linear coefficient $m$ and a constant offset to each two-dimensional distribution $\log{}R_{\rm{FP}}-\log{}R_{\rm{vir*/eff}}$, showing the results in green in Fig. \ref{fig:fittedFPs}, for comparison with $1:1$ relations, given as cyan lines.

The subsets of lower-mass objects in each Full FP are shown in Fig. \ref{fig:fittedFPs} as coloured contours, encompassing 68\% and 95\% of the supplemental objects, and labelled by the mass range of the subset: $\log\left(M_{*}/\msol\right)<10.5$. The minimum elliptical stellar mass is $10^{10}\msol$, whilst spirals in the Full FP sample go down to $\sim10^{8.5}\msol$ (see Fig. \ref{fig:distributions}). The FP root-mean-square residuals $\sigma_{\rm{FP}}$ are given as black and coloured numbers\footnote{These correspond to the rms of $\lambda$ divided by $\ln10$, for comparison with previous work \citep{joachimi,Singh2020,Rosito2020}.} per-panel, corresponding to the Full FP and the low-mass subset, repsectively.

Fitted FP coefficients are displayed in Table \ref{tab:sample_details}. The signs of the coefficients show that Masscut ellipticals with large radii have higher velocity dispersion and lower surface mass density. These trends are consistent with those seen for the Full FP sample, excepting a smaller intercept $c$, which is reflected in Figs. \ref{fig:distributions} \& \ref{fig:fittedFPs} as the low-mass subset can be seen to inhabit smaller radii.

Our spiral FP coefficients and intercept exhibit agreeable signs with those found by \cite{Shen2002}\footnote{Considering those of their FP definitions that are comparable with our own; their Eqs. 13 \& 14, Tables 1-3.}, who suggested negative values for each (if radii are taken in kiloparsecs). Both the magnitude and circular velocity slopes are significantly steeper for the Masscut FP sample than for the Full FP, whilst the intercept is slightly less negative. These differences reflect the large space of $M_r-V_{\rm{c}}$ opened-up by the addition of lower-mass galaxies.

We observe more of a tilt in the spiral plane, and a marginally larger rms scatter $\sigma_{\rm{FP}}$, though the spiral FP is qualitatively similar to that of ellipticals. We note also that the Full FP is less tilted than the Masscut FP for both ellipticals and spirals. These findings are promising for FP studies wishing to explore the evolution and statistics of deep, spiral-dominated galaxy samples.

A comparison with the previous study of the elliptical galaxy FP in Horizon-AGN \citep{Rosito2020} is not direct; as discussed in Sect. \ref{sec:simulation}, that analysis used differently defined three-dimensional/projected radii and surface areas. These, as well as minor selection differences, conspire to yield incompatible coefficients in the fit -- though they deviate in the same directions from the virial-theorem-constant-$M/L$ prediction ($a=2,b=-1$), and the rms residuals are comparable ($\sim0.071-0.074$ in this work, vs. $0.067$ in \citealt{Rosito2020}).

\begin{figure*}
    \centering
    \includegraphics[width=\textwidth]{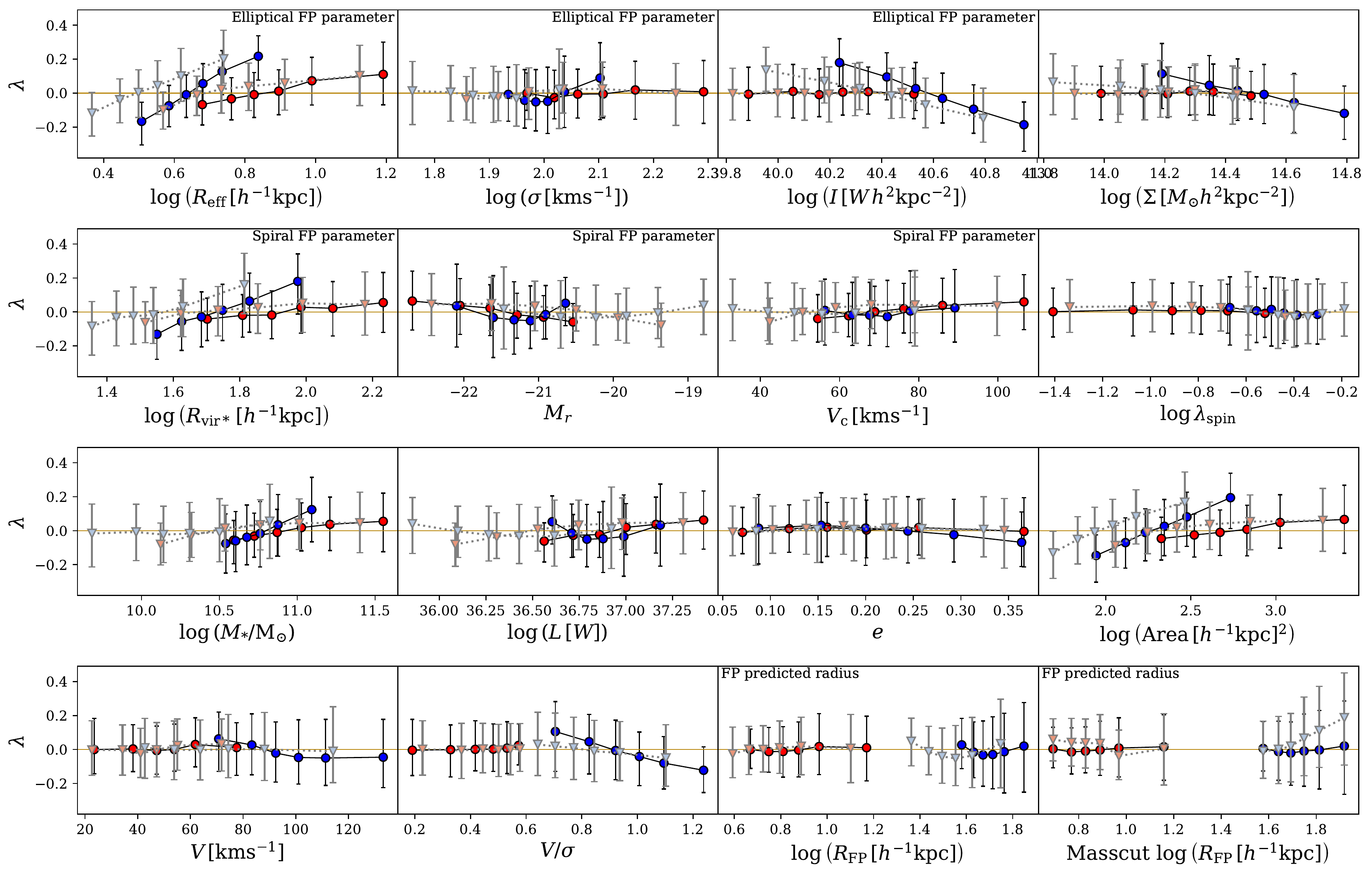}
    \caption{The intrinsic sizes $\lambda$ of spiral (blues) and elliptical (reds) galaxies, as estimated from Fundamental Plane residuals (Sect. \ref{sec:fundamental_planes}), vs. various galaxy properties (see Sect. \ref{sec:simulation} for details on these and other properties). $\lambda_{\rm{spin}}$ is the spin parameter \citep[][Eq. 2]{Aubert2004}, not to be confused with Fundamental Plane residuals $\lambda$. Shown are the relations both for conservative Masscut Fundamental Plane samples ($M_{*}>10^{10.5}\msol$; darker colours), and for Full Fundamental Plane samples (lighter colours; see Table \ref{tab:sample_details} for sample details). Points and error bars show the means and $16^{\rm{th}},84^{\rm{th}}$ percentiles of $\lambda$ for six equipopulated bins defined on each of the $x$-axes. Several galaxy property variables exhibit correlations with the Fundamental Plane residuals $\lambda$, which must be taken into account when considering their spatial correlations. Panel headings indicate in which cases the quantity in the $x$-axis is part of the Fundamental Plane fits. Deviations from the horizontal in this case indicate a residual dependency on the parameter that could not be captured by the best-fit Fundamental Plane. For panels without headings, deviations from the horizontal suggest that there is a correlation between sizes and the parameter considered, and that including it as a variable in the Fundamental Plane fit could improve the predictability of galaxy sizes.
    }
    \label{fig:lambda_vs_gxy_props}
\end{figure*}

Fig. \ref{fig:lambda_vs_gxy_props} displays relationships between our Masscut and Full FP residuals $\lambda$ and galaxy properties. Elliptical $\lambda$ exhibit a weakly positive correlation with the effective radius, as expected given the slight tilting of the plane. The softly positive trend of $\lambda$ with measured radii is similar to that seen by \cite{Singh2020}, who also saw strong negative correlations between $\lambda$ and absolute magnitude, which we do not observe. We observe slightly stronger positive correlations of spiral $\lambda$ with virial (due to tilting) and effective radii, and with stellar mass. Correlations between $\lambda$ and other FP variables are weak, barring some correlations of spiral $\lambda$ with projected surface quantities -- none of which feature on the spiral FP -- and an anti-correlation with the morphological metric $V/\sigma$. Thus faster-spinning spirals tend to be intrinsically smaller.

The Masscut FPs are qualitatively similar to the Full FPs, but featuring slightly stronger tilts in both cases (though within $1\sigma$ of the gradients quoted in Fig. \ref{fig:fittedFPs}), and losing significant numbers of low-radius objects. Differences between $\lambda$ estimates with and without the mass selection are uncorrelated with stellar mass.

We note that the mass-selection tends to exacerbate correlations between FP residuals and other galaxy properties (dark vs. light points in Fig. \ref{fig:lambda_vs_gxy_props}) in all cases -- thus Full FP residuals are less contaminated by object characteristics other than intrinsic size fluctuations. We take these various correlations into consideration when interpreting measured ISCs in the coming sections.

Our Horizon-AGN samples thus yield good fits of the FP, with elliptical coefficients and rms scatter that are comparable with those found for observational data \citep{Hyde2009,Saulder2013,joachimi,Saulder2019,Singh2020a} and hydrodynamical simulations (Horizon-AGN; \citealt{Rosito2020}, IllustrisTNG; \citealt{Lu2020}). For simulated spiral galaxies, the planes are comparable with the theoretical planes of \cite{Shen2002}, and feature stronger tilts than the elliptical planes. Still, correlations between FP residuals and key properties, such as the virial radius and stellar mass, are not drastic -- particularly in the case of the Full FP. We thus advance to the measurement of intrinsic size correlations so-derived, applying caution as regards the impacts of tilting, $\lambda$-property correlations, and of our usage of virial radii to characterise spiral sizes.

%--------------------------------------------------------------------

\section{Intrinsic size correlations}
\label{sec:intrinsic_size_correlations}

We follow \cite{joachimi} and \cite{Singh2020} in measuring galaxy clustering and intrinsic size correlations, though we do not integrate signals along the line-of-sight (the simulation box $z$-axis in our work), preferring to leverage our exact knowledge of object positions to assess correlations as functions of three-dimensional pair separations. The relevant correlation functions are estimated via
\begin{equation}
    \hat{\xi}_{\rm{gg}}(r)=\frac{DD-2DR_D+R_DR_D}{R_DR_D} \label{eq:lse} ,
\end{equation}
\begin{equation}
    \hat{\xi}_{\rm{g}\lambda}(r)=\frac{D\lambda-R_D\lambda}{R_\lambda R_D}  ,
    \label{eq:glambda}
\end{equation}
\begin{equation}
    \hat{\xi}_{\lambda_1\lambda_2}(r)=\frac{\lambda_1\lambda_2}{R_{\lambda_1} R_{\lambda_2}} \label{eq:twolambdas}  ,
\end{equation}
 where $DD,RD,RR$ are unweighted pair-counts between galaxy $D$ and random point $R$ samples (with subscripts denoting galaxy samples to which the randoms correspond), for measurements of galaxy clustering. These, and the weighted pair-counts $D\lambda,R_D\lambda,\lambda_1\lambda_2$, are given by
\begin{equation}
    XY = \sum_{i,j} \Theta_{ij}, \\
\end{equation}
\begin{equation}
    X\lambda = \sum_{i,j} \lambda_{i}\,\Theta_{ij}, \\
\end{equation}
\begin{equation}
    \lambda_1\lambda_2 = \sum_{i,j} \lambda_{1,i}\lambda_{2,j}\,\Theta_{ij} ,
\end{equation}
where samples $X,Y$ denote the galaxy $D$ or randoms $R,R_{D},R_\lambda$ samples, $i,j$ are indices running over all objects in the two samples, and $\Theta_{ij}$ are binary filters applying a log-spaced binning in three-dimensional separation $r$. Notice Eq. \ref{eq:lse} is the usual Landy-Szalay estimator \citep{Landy1993}. All estimators presented in Eqs. \ref{eq:lse}, \ref{eq:glambda} and \ref{eq:twolambdas} are normalized by random-random pair counts (vs. data-data and data-random pair counts). This is to allow direct comparison of our results to previous works \citep{joachimi,Singh2020}, and to ensure that they can be interpreted and modelled as correlation functions in the sections that follow.

We measure the auto-clustering correlations $\hat{\xi}_{\rm{gg}}$ for each of our defined FP samples (Table \ref{tab:sample_details}), and also split those samples into intrinsically large $\lambda_+$ and small $\lambda_-$ subsets for assessment of the clustering variation across the FP. 

We define two additional `lens' samples of elliptical galaxies, with masses $>10^{10}\msol$ and $>10^{10.5}\msol$ -- the latter, more conservative mass selection matching that applied to our Masscut FP samples (see Sect. \ref{sec:simulation}) -- for use as density tracers in the density-size cross-correlations $\hat{\xi}_{\rm{g}\lambda}$. The lens samples are similar to the elliptical FP samples, but retain galaxies lost to poor fitting of the FPs, thus offering slight reductions in shot-noise. Intrinsic size auto-correlations $\hat{\xi}_{\lambda\lambda}$ are measured only within FP samples, since $\lambda$ probes different radii between ellipticals and spirals (Sect. \ref{sec:simulation}), and is derived from differently fitted and tilted planes between the Masscut and Full setups.

We note that, whilst our fitted $\lambda$ distributions are extremely close to symmetric by construction, any $\left\langle\lambda\right\rangle\neq0$ will result in scale-invariant additive contributions to $\lambda$-weighted correlation functions, most visible at large separations where correlations are weak, and stemming from inherent clustering contributions \citep{Singh2020}. We ensure that this contribution is compensated by subtracting $\left\langle\lambda\right\rangle$ from each $\lambda$ distribution under consideration, though in practice, for these samples, this subtraction is small and not necessary for the measurement of stable signals in the simulation box.

We work with 7 three-dimensional separation bins in the range $0.5-30\mpch$, and estimate covariances according to a delete-one jackknife resampling of 64 subvolumes defined on the simulation box. The covariance estimates are given by
\begin{equation}
    \widehat{\rm{Cov}} = \frac{63}{64}\sum^{64}_{i=1}\left(\xi_i - \bar{\xi}\right)^{\rm{T}}\left(\xi_i - \bar{\xi}\right)\,,
    \label{eq:jackknife_covariance}
\end{equation}
where $\xi_i$ is a correlation function measured upon removal of the $i$'th jackknife subsample, $\bar{\xi}$ is the average of all jackknife measurements, and $\rm{T}$ denotes the conjugate transpose of the difference vector.

We exclude the final data-point from assessments of signal significance, since the variance on these scales is unlikely to be captured by subvolumes of scales $\sim{}L_{\rm{box}}/4=25\mpch$. The resulting Hartlap correction factor \citep{Hartlap2007} for estimates of the 6-bin signal inverse covariance is then $\sim0.9$, and the maximum scale under consideration is $\sim17\mpch$.

FP size residuals $\lambda$ are not re-normalised, so as to preserve the amplitude of the signal, which otherwise would be inconsistent with the estimation of contamination to magnification signals. To explore the symmetry of intrinsic size correlation signals, we also measured various correlations for absolute sizes $|\lambda|$, and for sample selections $\lambda_+$ and $\lambda_-$, finding that most such splittings yielded signals corresponding to the intrinsic clustering of objects, merely down-weighted by powers of $\meanl$\footnote{One could re-subtract $\meanl$ from the split distributions, but the meaning of the FP residual would then be lost; we neglect to explore the split-$\lambda$ signals further, noting that any interesting asymmetries in the behaviour of $\lambda_+,\lambda_-$ galaxies should manifest as non-zero signals measured on the original $\lambda$ distributions.}.

The exception to this statement is for galaxy clustering measured within $\lambda_+$ and $\lambda_-$ samples, which shows variably significant differences for spiral and elliptical samples; meaning that, depending on the galaxy morphology, intrinsically large/small objects cluster differently in the simulation. We follow \cite{joachimi} in defining a statistic to capture this variability, given as
\begin{equation}
    \dgg(r) = \frac{\xigg(r,\lambda>0)}{\xigg(r,\lambda<0)} - 1 \, .
    \label{eq:delta_gg}
\end{equation}
Retaining individual clustering measurements from jackknife resampling, we are able to estimate the covariance of this quantity directly, thus achieving a partial cancellation of shot-noise and cosmic variance.

Random catalogues are oversampled ($20\times$ relative to their corresponding galaxy sample, given by subscripts on $R$) sets of points, uniform-randomly distributed in the simulation box with unit weights. All correlations and jackknife covariances are implemented within {\sc{TreeCorr}} \citep{Jarvis2013}, taking advantage of the functionality designed for scalar convergence fields $\kappa$, and observing periodic boundaries of the simulation box during pair-counting of three-dimensional correlations.

\subsection{Lensing contamination}

We focus now on
ways in which ISCs might contaminate galaxy statistics
by mimicking the size variations induced by weak lensing magnification,
in a near-exact analogy to the intrinsic alignment phenomenon that contaminates studies of cosmic shear.

\subsubsection{Density-magnification}
\label{sec:density_magnification_contamination}

\citet{Huff2011} give an estimator for the weak lensing convergence $\hat\kappa$ as
\begin{equation}
    \log(1+\hat\kappa)\equiv\Delta\log R \,,
    \label{eq:huffgraves_kappa}
\end{equation}
where $\Delta\log{R}=\log{}R_{\rm{O}}-\log{}R_{\rm{FP}}$ is the residual between an observed galaxy radius $\log{}R_{\rm{O}}$ and an FP prediction thereof, given by $\log{}R_{\rm{FP}}$. Cross-correlating this quantity with foreground lenses, as demonstrated by \cite{Huff2011}, one can construct a two-point estimate of the projected surface density $\Sigma$ at lens plane $z_l$, via
\begin{equation}
    \Sigma_{\rm{crit}}\kappa = \Sigma(d_l\vec{\theta})\,,
    \label{eq:sigma_r}
\end{equation}
for the comoving distance to the lens plane $d_l$, and angular separation vector $\vec{\theta}$, with critical surface density
\begin{equation}
    \Sigma_{\rm{crit}} = \frac{c^2}{4\pi{}G}\frac{d_{\rm{a}}(z_s)}{d_{\rm{a}}(z_l)\left(d_{\rm{a}}(z_s)-d_{\rm{a}}(z_l)\right)(1+z_l)^2}\,,
    \label{eq:sigma_crit}
\end{equation}
where $d_{\rm{a}}$ denote angular diameter distances, and subscripts $s,l$ denote source (background) and lens (foreground) samples.

Intrinsic size fluctuations will contribute to the estimate $\hat\kappa$ \citep{Alsing2015,Ciarlariello2015,Ciarlariello2016}. If these fluctuations do not vanish under spatial averaging, e.g. if the distribution is asymmetric, or if they are correlated with one another or with structure, then this may introduce biases into estimators such as Eq. \ref{eq:sigma_r} via some induced deviation from the true mean size of objects at a given redshift.

We return to Eq. \ref{sec:observed_radius_expanded}, which re-defines the observed radius $R_{\rm{O}}$ in terms of the lensing convergence $\kappa$, the FP-predicted radius $R_{\rm{FP}}$, and the lensing-independent intrinsic size fluctuation $\lambda$, with $\meanl\rightarrow0$, i.e. the global distribution of intrinsic galaxy sizes is symmetric. This is an assumption, but one supported by the fitted low-$z$ (and therefore low-lensing) FP of \cite{joachimi}, and borne out by our own simulation data\footnote{We might expect a violation of this symmetry in deep lensing data; at high redshifts, shape measurements necessitate resolution cuts that will exclude the small-$R$ tail of objects, even if the true $\lambda$ distribution is symmetric. This is the source of the lensing and size biases \citep{Schmidt2009a,Schmidt2009,Ciarlariello2015,Ciarlariello2016}, as well as an intrinsic size selection bias in the presence of ISCs.}. We then expand Eq. \ref{eq:huffgraves_kappa} as
\begin{equation}
    \log\left\{\left(1+\kappa\right)\left(1+\lambda\right)\right\} = \log\frac{R_{\rm{O}}}{R_{\rm{FP}}}\,.
    \label{eq:redefine_RI}
\end{equation}
We note that $\lambda$ here (and in Eq. \ref{sec:observed_radius_expanded}) is approximately equivalent to our definition in Eq. \ref{eq:deflambda}, modulo terms of $\mathcal{O}\left(\lambda^2\right)$ \citep{joachimi}. The estimator of Eq. \ref{eq:huffgraves_kappa} then becomes
\begin{equation}
    \Delta\log{}R \equiv \log\left(1+\widehat{\kappa+\lambda}\right)\,,
    \label{eq:estimator_wISC}
\end{equation}
where $\widehat{\kappa+\lambda}$ signifies that the \cite{Huff2011} estimator is actually estimating the combination of intrinsic and lensing-derived size fluctuations. If intrinsic sizes are uncorrelated with structure, then one expects $\lambda$ to vanish under spatial averaging, and the unbiased convergence estimate of Eq. \ref{eq:huffgraves_kappa} is recovered. This would not, however, preclude the possibility of auto-correlations in $\lambda$, which would have the potential to contaminate any magnification-magnification correlation estimates based on measured galaxy sizes (Sect. \ref{sec:magnification_magnification_contamination}).

An ISC contamination of estimates for $\Sigma(r)$ may be of lesser concern, as one expects that lens and source samples should not feature ISCs if the constituent galaxies are not co-located in space; background intrinsic sizes know nothing about distant, foreground lenses. In analogy to the contamination of galaxy-galaxy lensing (GGL) by intrinsic alignments, one would only expect ISC contamination of $\Sigma(r)$ in the presence of large photometric redshift errors -- however, such errors cannot be discounted, and are known to be more prevalent among the spiral galaxies that dominate weak lensing samples \citep{Rozo2016}.

In \cite{Huff2011}, angular cross-correlations $w_{ls}(\vartheta)$ between lens and source samples were used to estimate the fraction $f_l$ of foreground sources that had been scattered out to higher redshifts by photo-$z$ errors. These galaxies acquire systematically biased FP residuals, as they are thought to be far more distant than they are in reality. The interloper fraction $f_l\,(z_l,z_s,\vartheta)\in[0,1]$, per lens and source redshift $z_l$ and $z_s$, and angular $\vartheta$ bin, was thus used to down-weight the convergence estimate as $(1-f_l)\kappa$; to weight a term in the estimator $f_l\Delta\log{R}_{\rm{err}}$ that corrects the FP residual of a redshift interloper; and to outright discard lens-source bin pairs and angular scales for which the interloper fraction is very high.

In our simulated data, the convergence $\kappa=0$. Thus we can estimate the amplitude of an ISC contribution $\Sigma_\lambda(r)$ by cross-correlating FP residuals $\Delta\log{}R$ with galaxy positions, assuming some distribution of photometric redshift errors. The hypothetical setup is that galaxies situated at some low redshift have been mistaken for higher-redshift sources, and thus made their way into an estimate of the projected surface density. Assuming some value or functional form for $f_l$, which would in reality be estimated/modelled using data/simulations, we can also approximate the mitigation strategy of \cite{Huff2011}.

We note that $f_l$ could in principle vary widely between zero and unity, though all efforts will seek to minimise it. For a single $\vartheta$-bin $\in[0.6,6]\,\rm{arcmin}$\footnote{Corresponding to scales of order a few hundred $\kpch$ at $z\sim0.1$, and a few/several $\mpch$ at $z\sim2$, which are coincident with those observed by \cite{Huff2011}.}, \cite{Stolzner2022} estimate the lens-source vs. lens-lens angular clustering ratio $w_{ls}/w_{ll}$ ($\neq{}f_l$) to increase with redshift, and to be as large as $\sim0.4$ ($\sim0.1$), for spectroscopic lenses, without (with) an outlier mitigation strategy for the LSST DESC cosmoDC2 mock catalogue \citep{Korytov2019}. One expects larger ratios for photometric lens samples, as redshift errors should deprecate the auto-correlation $w_{ll}$ more severely than the cross-correlation $w_{ls}$. These ratios are not equivalent to $f_l$ -- the fraction of `source' objects that are in fact situated at the lens plane -- but they do indicate that the strength of spurious, outlier-driven clustering is expected to remain relatively high for the foreseeable future. Noting also that \cite{Huff2011} excluded any redshift-angular separation bins for which $f_l>0.5$ (and found higher thresholds to change little), we explore a simple grid formed of three points in $f_l$ at $0.0,0.2,0.4$, and a standalone case of $f_l=1$, which is the true case for our hypothetical setup; all of our `source' objects are redshift interlopers, co-located with the lenses.

For simplicity, we henceforth assume insignificant redshift evolution in $f_l$, and in the intrinsic size field, and its correlations with itself and the density field. Thus we treat the box as if it were situated at redshifts up to $z\sim0.2$ (and extrapolate model fits at $z=0.06$ to $z\sim0.8$ in Sect. \ref{sec:magnification_magnification_contamination}) and make predictions for ISC lensing contamination with the caveat that evolution of the intrinsic size field, cross-talking with $z$-dependent photo-$z$ quality, will pose further complications for a magnification analysis\footnote{We note also that, if ISCs are in fact stronger at higher-$z$, then our estimates of contamination are conservative. If they are weaker at high-$z$, then their amplitude must have grown over time. Concurrently, the Universal star formation rate, and elliptical galaxy fraction, are evolving, particularly as observed in flux-limited samples contending with the Malmquist bias. The relative strength of spiral/elliptical ISCs would then be of interest with respect to the evolution of the contamination with redshift, similarly to cosmic shear contamination by IA \citep{Fortuna2020}.}. Since we retain lower-redshift FP residuals when scattering objects out to $z_s$, the part of the FP error term $\Delta\log{R}_{\rm{err}}$ that corrects $R_{\rm{FP}}(z_s)\rightarrow{}R_{\rm{FP}}(z_l)$ is effectively already applied, and without any error. We thus mimic the correction term simply according to the ratio of angular diameter distances $d_{\rm{a}}$ at the mistaken source redshift and true lens redshift \citep{Huff2011}
\begin{equation}
\begin{split}
    \Delta\log{R}_{\rm{err}}(z_l,z_s) =& \log\left(\cfrac{d_{\rm{a}}(z_s)\,R_{\rm{FP}}(z_l)}{d_{\rm{a}}(z_l)\,R_{\rm{FP}}(z_s)}\right) \\ \rightarrow& \log\left(\cfrac{d_{\rm{a}}(z_s)}{d_{\rm{a}}(z_l)}\right) \,.
    \label{eq:FP_error_term}
\end{split}
\end{equation}
The final intrinsic size field estimator, featuring the photo-$z$ correction term, then becomes
\begin{equation}
    \Delta\log{R} = \log{\left(1+\hat\lambda\right)} + f_l\,\Delta\log{R}_{\rm{err}}(z_l,z_s)\,.
    \label{eq:estimator_lambdahat}
\end{equation}
In order to estimate the intrinsic contribution, we translate the centre of the simulation box to a fixed comoving distance $\chi(z_l=0.2)$, and convert galaxy three-dimensional Cartesian coordinates into RA, Dec, and $\chi$. Size tracer objects are thus situated at $z\sim0.2$, with corresponding RA, Dec, but we shall mistake them to be at some higher $z_s$, whilst all lenses are at $z_l\sim0.2$. This means that $\Sigma_\lambda(r)$ is most likely to feature at large-$r$, since small-angle lens-source pairs will be thought to probe highly non-linear physical scales on the lens plane that are difficult to model, and often excluded from GGL analyses \citep[see][]{Singh2020a}.

As mock lenses, we employ the Lens samples described in Sect. \ref{sec:intrinsic_size_correlations} (Table \ref{tab:sample_details}). For source galaxies, \cite{Huff2011} selected 8.4 million photometric SDSS \citep{York2000} elliptical galaxies between $z=0.1-0.55$. We fit an analytic $n(z)$ to the reconstructed redshift distribution of the SDSS DR8 photometric sample \citep{Sheldon2012}, given by
\begin{equation}
    n(z) = z^\alpha \exp\left\{-\left(\frac{z\sqrt{2}}{z_0}\right)^{\beta}\right\}\,,
    \label{eq:nz_fit_sheldon}
\end{equation}
with $\alpha=1.58,\,z_0=0.34,\,\beta=1.86$. Assuming that our measurable density-FP residual correlations are representative across the redshift range of interest, we bootstrap our FP samples to estimate the uncertainty in $\Sigma_\lambda(r)$, neglecting cosmic variance; our predictions might thus be considered pessimistic. Each FP sample (Table \ref{tab:sample_details}) is re-sampled with replacement 1000 times, and the bootstrapped galaxies\footnote{We note that the elliptical samples are subsets of the lens samples given in Table \ref{tab:sample_details}, and would thus result in unrealistic, duplicated galaxy positions between foreground lenses and background sources. We verify that discarding lens objects that are duplicated in the source sample prior to measuring each of the bootstrap correlations has a negligible impact upon the signal prediction other than to raise the noise level.} are assigned redshifts $z_s$ drawn from the fitted $n(z)$ (Eq. \ref{eq:nz_fit_sheldon}). We note that the hard $f_l<0.5$ cut imposed by \cite{Huff2011} functions to exclude regimes dominated by source galaxies that are physically coincident with lenses; we thus exclude galaxies assigned to $0.1<z_s<0.3$, i.e. we impose that no $z_s$ sits within $z_l\pm0.1$. We deliberately allow a small number of $z_s<0.1$ objects, but find that they make no discernible difference to the measured correlation.

We estimate $\Sigma_\lambda(r)$ itself by cross-correlating the intrinsic size field estimates $\hat\lambda(f_l,z_s)$ (Eq. \ref{eq:estimator_lambdahat}), weighted by critical surface densities $\Sigma_{\rm{crit}}(z_l,z_s)$ (Eq. \ref{eq:huffgraves_kappa}), with angular lens sample positions, collecting the weighted-average intrinsic size residual $\langle\hat\lambda\rangle(\vartheta)$ that would contaminate\footnote{This contamination will not function via simple addition, since the denominator in our estimator includes only the interloper objects; $\Sigma_\lambda(r)$ will contribute to any measured $\Sigma(r)$ only after a re-weighting according to the relative numbers of lens-source and lens-interloper pairs in each angular bin.} the convergence estimate $\langle\hat\kappa\rangle(\vartheta)$. The contamination estimator is thus given as
\begin{equation}
    \hat\Sigma_\lambda(\vartheta) = \cfrac{\sum_{i,j} \Sigma_{\rm{crit}}(z_i,z_j) \, \hat\lambda_j\, \langle{}i|j\rangle}{\sum_{i,j} \Sigma_{\rm{crit}}(z_i,z_j)}\,,
    \label{eq:Sigma_lambda_estimator}
\end{equation}
where $i,j$ index lens and source samples, respectively, $\hat\lambda_j$ gives the intrinsic size ($\hat\kappa$-contaminant) estimate for galaxy $j$ (Eq. \ref{eq:estimator_lambdahat}), and the bin-filter $\langle{}i|j\rangle$ here applies the angular binning $\vartheta$. Angular separations $\vartheta$ are then converted into transverse comoving separations\footnote{Presented in kpc, and not $\kpch$, for more direct comparison with \cite{Huff2011}.} $r$, all evaluated at $z_l=0.2$. We further measure the average size residual around randomly distributed points, for subtraction from the galaxy signal; this removes spurious large-scale signals, and improves the covariance properties of the estimator \citep{Singh2016}. 

As for the box correlations (Sect. \ref{sec:intrinsic_size_correlations}), here one must consider the mean FP residual $\meanl$ which can contribute spurious signals derived from the clustering of galaxies. Observational galaxy FP analyses incorporate redshift dependence via polynomial fitting \citep{joachimi,Singh2020}, or else fit the FP in narrow redshift bins \citep{Huff2011,Singh2020}, thus achieving $\meanl\sim0$ over the redshift range, or enforcing it via subtraction. However, selections and effective weighting in $f_l$ could reintroduce a non-zero mean residual, if imposed after fitting of the FP. We assess the impact of the mean residual for our setup by re-measuring $\Sigma_\lambda(r)$ after enforcing $\meanl=0$, and find that the difference in signal is negligible due to our subtraction of the signal measured around random points. The random signal subtraction is itself necessary to remove a spurious excess signal at large $r$, even if $\meanl=0$ is enforced.

For these angular correlations, we avoid a consideration of periodic boundaries by limiting the lens sample to be at least $5\mpch$ away from the inner-edge of the box on the $X-Y$ plane. At the lens redshift of $z_l=0.2$, this corresponds to an angle of $\sim1.7\,\rm{arcmin}$, beyond which we discard measured correlations. \cite{Huff2011} estimated the effect of photo-$z$-induced mis-estimation of $\Sigma_{\rm{crit}}$ upon their signal to be less than 10\%, which should hold for our work (modulo noise) as we adopt their redshift distribution and consequent lensing geometry -- as we shall see, the $f_l$-weighted correction term, and in particular our mass-selections, wield far larger influence over the signals in any case.

\subsubsection{Magnification-magnification}
\label{sec:magnification_magnification_contamination}

We also consider the potential for ISCs to contaminate estimates of magnification auto-correlations $\kappa\kappa$. Again in analogy to the phenomena of IA, a density-intrinsic size correlation $\delta\lambda$ could induce spurious contributions to size-based $\kappa\kappa$ estimates over large separations in redshift, as the same lenses source both background convergence and foreground ISCs -- the gravitational-intrinsic (GI) correlation, in IA parlance \citep{Hirata2004a} -- whilst the impact of auto-correlations $\lala$ would be largely limited to tomographic auto-correlations which include many closely-associated galaxy pairs -- the intrinsic-intrinsic (II) correlation.

In tidal torque theory \citep{Schaefer2008}, intrinsic spiral galaxy ellipticities and alignments are expected to be pure orientation effects that will not yield a GI correlation \citep{Tugendhat2020,Ghosh2020}. However, this null prediction will not necessarily hold for intrinsic spiral sizes, because the size relates to the isotropic part (i.e. the trace) of the tidal shear tensor; itself proportional to the local density via Poisson's equation \citep{Tugendhat2018}. Thus one should expect a GI magnification correlation from spirals, if indeed spiral sizes are correlated with the local density contrast $\delta$.

Moreover, one might expect a significant spiral intrinsic size auto-correlation to contribute strongly to redshift bin auto-correlations $\kappa\kappa$ as an II term, since spiral galaxies dominate deep lensing data; a fact that also makes exclusion of such objects difficult, given the deleterious impact upon the signal-to-noise of the desired measurement.

Lacking measurements of $\kappa\kappa$ correlations in the literature with which to compare, we elect instead to make a similar ansatz to those made by \cite{joachimi} and \cite{Alsing2015}\footnote{Though \cite{Alsing2015} modelled intrinsic and magnification-induced size variations according to joint distributions of absolute size and flux, as opposed to deviations from the FP.}; that the intrinsic size field $\lambda(\vec{x})$ can be characterised by some linear scaling of the matter density contrast $\delta(\vec{x})$, parameterised separately for spiral and elliptical sizes as $B_{\rm{spi}}$ and $B_{\rm{ell}}$, respectively.

We fit this model to our measured $\lala$ correlations in the range $3<r<17\mpch$, where the lower limit excludes highly non-linear scales, whilst the upper is dictated by our sub-sample covariance estimation procedure (see Sect. \ref{sec:intrinsic_size_correlations}). We thus constrain the absolute values of the $B$ parameters, the squares of which modulate the amplitude under our ansatz: $\xill=B^{2}\xi_{\delta\delta}$, where $\xi_{\delta\delta}$ is the matter auto-correlation function, computed for the Horizon-AGN cosmology using {\sc{nbodykit}}\footnote{\url{https://nbodykit.readthedocs.io}} \citep{Hand2018}.

For comparison with \cite{joachimi}, we assume two simple Gaussian redshift distributions centred on $z=0.4,0.8$ with widths $\sigma=0.1$ -- we thus assume the phenomenological model from $z=0.06$ to hold out to $z\sim0.8$, which is clearly optimistic. However, our predictions should be considered conservative in the case that ISCs weaken over cosmic time, and as we shall see, the landscape of spiral/elliptical predictions makes for interesting conclusions in any case. We convert the model into $C_{\ell}$ expectations for the auto- $\lala$ and cross-correlations $\lambda\kappa$, to be compared with the expected convergence signal $\kappa\kappa$ via \citep{joachimi}
\begin{eqnarray}
    C^{ij}_{\kappa\kappa}(\ell) &=& \int^{\chi_{\mathrm{hor}}}_{0} \, \mathrm{d}\chi\, \cfrac{q^i(\chi)\,q^j(\chi)}{\chi^2}\,P_{\delta}\left(\cfrac{\ell+1/2}{\chi},\,\chi\right)\,, \nonumber \\
    C^{ij}_{\lambda\kappa}(\ell) &=& B\: \int^{\chi_{\mathrm{hor}}}_{0} \, \mathrm{d}\chi\, \cfrac{p^i(\chi)\,q^j(\chi)}{\chi^2}\,P_{\delta}\left(\cfrac{\ell+1/2}{\chi},\,\chi\right)\,, \label{eq:Cells} \\
    C^{ij}_{\lala}(\ell) &=& B^2\: \int^{\chi_{\mathrm{hor}}}_{0} \, \mathrm{d}\chi\, \cfrac{p^i(\chi)\,p^j(\chi)}{\chi^2}\,P_{\delta}\left(\cfrac{\ell+1/2}{\chi},\,\chi\right)\,, \nonumber
\end{eqnarray}
where we take $P_\delta$ as the matter power spectrum with non-linear corrections\footnote{We note that \cite{Alsing2015} used the linear power spectrum for II and the geometric mean of non-/linear spectra for GI ISCs, as suggested by \cite{Kirk2012} in the context of IA. Recent years have seen declining usage of this model in IA contexts, and we elect to use the full non-linear power spectrum here.} \citep{Smith2003,Takahashi2012}, $p^{i}(\chi)$ is the probability distribution of comoving distances $\chi(z)$ in the $i$'th tomographic sample, $\chi_{\rm{hor}}$ is the comoving distance to the horizon, $\ell$ is the angular multipole, $B$ is the relevant linear intrinsic size field parameter, and $q^{i}(\chi)$ is the lensing efficiency of sample $i$, given by
\begin{equation}
    q(\chi) = \cfrac{3H^{2}_{0}\Omega_{\rm{m}}}{2c^2} \cfrac{\chi}{a(\chi)} \int_{\chi}^{\chi_{\rm{hor}}} \mathrm{d}\chi'\,p(\chi')\,\cfrac{\chi'-\chi}{\chi'}\,,
\end{equation}
assuming a flat universe, with present-day Hubble parameter $H_{0}$ and matter energy-density fraction $\Omega_{\rm{m}}$, scale factor $a$, and speed of light $c$.

We note here that a physically-motivated model for these correlations is highly desirable both for prospective studies of cosmic convergence, and for gaining insight into galaxy evolution from intrinsic sizes. For example, \cite{Ghosh2020} construct a unified linear model of elliptical galaxy intrinsic alignments and size correlations, asserting that the `elasticity' of ellipticals -- the constant of proportionality $D$ between observed shapes/sizes and the magnitude and orientation of the tidal shear \citep{Tugendhat2018} -- is responsible for both.

A more complex and promising avenue is the effective field theory (EFT) model of \cite{Vlah2019}, which is capable of modelling scalar biased tracers -- such as galaxy sizes, and other properties -- in principle accounting for all contributions up to a chosen order. Yielding coefficients describing the strength of contributions from all possible field combinations, insights into the evolution of galaxy sizes could be readily derived. An application of this model to ISCs, in an extended simulation analysis, could thus be of great value.

%--------------------------------------------------------------------

\section{Results \& discussion}
\label{sec:discussion}

\begin{figure}
    \centering
    \includegraphics[width=\columnwidth]{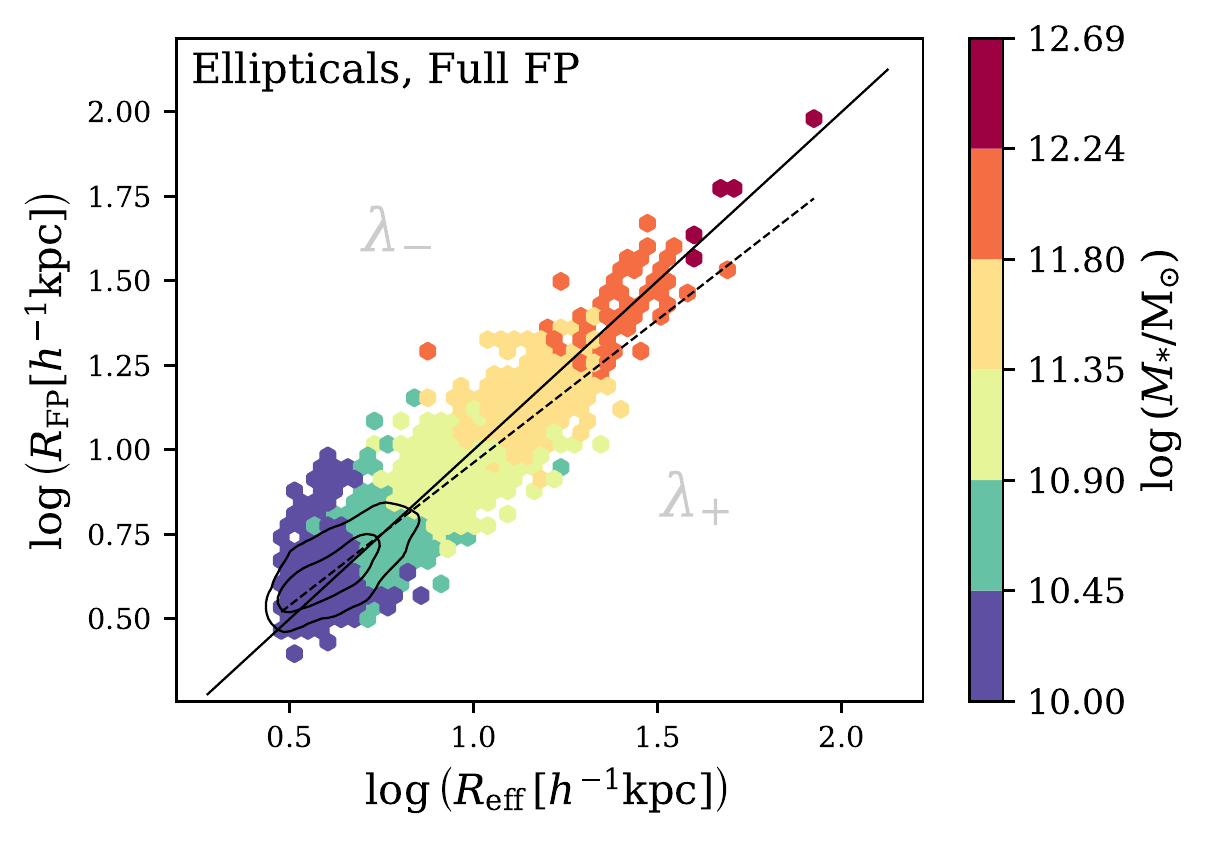}
    \caption{The Full elliptical sample Fundamental Plane (as in top-left panel of Fig. \ref{fig:fittedFPs}), coloured according to six uniform bins in log-stellar mass $\log{}(M_{*}/\msol)$. The $1:1$ relation, a linear fit to all points on the plane (Sect. \ref{sec:fundamental_planes_results}), and contours illustrating $M_{*}<10^{10.5}\msol$ objects (Sect. \ref{sec:simulation}), are reproduced from Fig. \ref{fig:fittedFPs}, here in black. Annotations $\lambda_-$ and $\lambda_+$ denote the sides of the plane corresponding to intrinsically small and large objects, respectively. Selections in stellar mass are seen to isolate regions of the Fundamental Plane, and similar trends are seen for spiral planes.
    }
    \label{fig:FPs_by_stellar_mass}
\end{figure}

Here we detail the results of our measurements of intrinsic size correlations, and our so-derived predictions for contamination of cosmic convergence statistics. We emphasise here the difference between absolute size and intrinsic size as we have defined it (Sect. \ref{sec:intrinsic_sizes}), noting that an absolutely large object can be intrinsically small (towards the right of a panel but above the cyan line, in Fig. \ref{fig:fittedFPs}) in comparison with its fellows of similar morphology and characteristics. We also note that many previous works explore the variability of galaxy sizes at fixed stellar mass. As Fig. \ref{fig:FPs_by_stellar_mass} shows, this is similar to considering the variability of FP residuals within an interval centred on some radius; that is, examining samples at fixed stellar mass is roughly akin to examining distinct regions of the FP.

\subsection{Intrinsic size correlations}
\label{sec:intrinsic_size_correlations_results}

\begin{figure*}
    \centering
    \includegraphics[width=\textwidth]{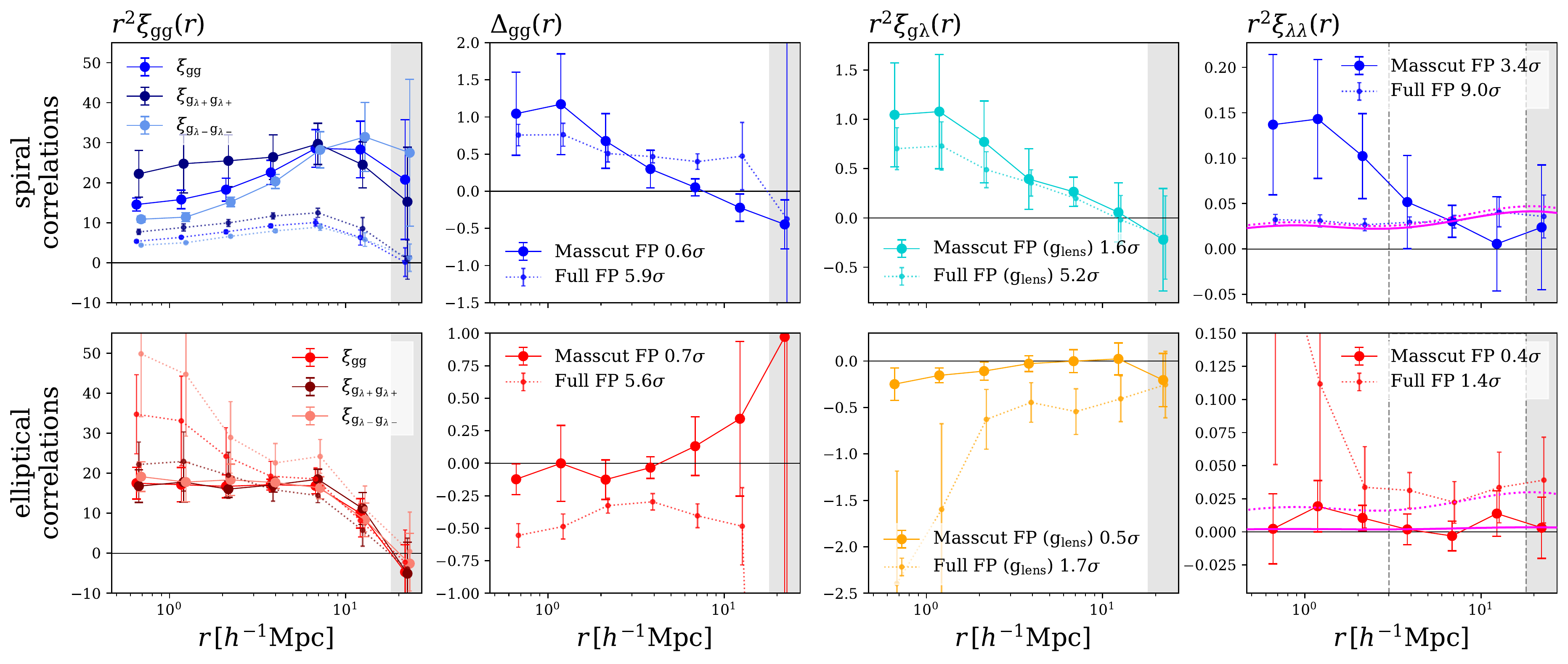}
    \caption{Galaxy two-point correlations, measured as functions of three-dimensional separations $r$, for Fundamental Plane (Sect. \ref{sec:fundamental_planes}) samples selected from the Horizon-AGN simulation (Sect. \ref{sec:simulation}). For spiral (\emph{top}) and elliptical (\emph{bottom}) galaxies, columns give the galaxy clustering $\xigg(r)$, the fractional difference in clustering $\dgg(r)$ between intrinsically large/small galaxies, the density-intrinsic size cross-correlation $\xigl(r)$, and the intrinsic size auto-correlation $\xill(r)$, respectively. The clustering panels also display the large (small) galaxy clustering signals $\xi_{\rm{g}_\lambda{+}\rm{g}_\lambda{+}}$ ($\xi_{\rm{g}_\lambda{-}\rm{g}_\lambda{-}}$) from which $\dgg$ is derived (Eq. \ref{eq:delta_gg}). 
    Larger data points denote more conservative signals measured for Masscut Fundamental Plane samples ($M_{*}>10^{10.5}\msol$; Sect. \ref{sec:simulation}), whilst smaller points are measured using the Full Fundamental Plane samples featuring lower-mass galaxies. Pink solid (dotted) curves in the right-most panels give the results of fitting the 1-parameter phenomenological model (described in Sect. \ref{sec:magnification_magnification_contamination}) separately to elliptical/spiral $\lala$ correlations measured in the Masscut (Full) Fundamental Plane samples. Intrinsic size correlations are highly sensitive to galaxies' morphology and stellar mass range.
    }
    \label{fig:ISCs}
\end{figure*}

\begin{figure*}[h]
    \centering
    \includegraphics[width=0.8\textwidth]{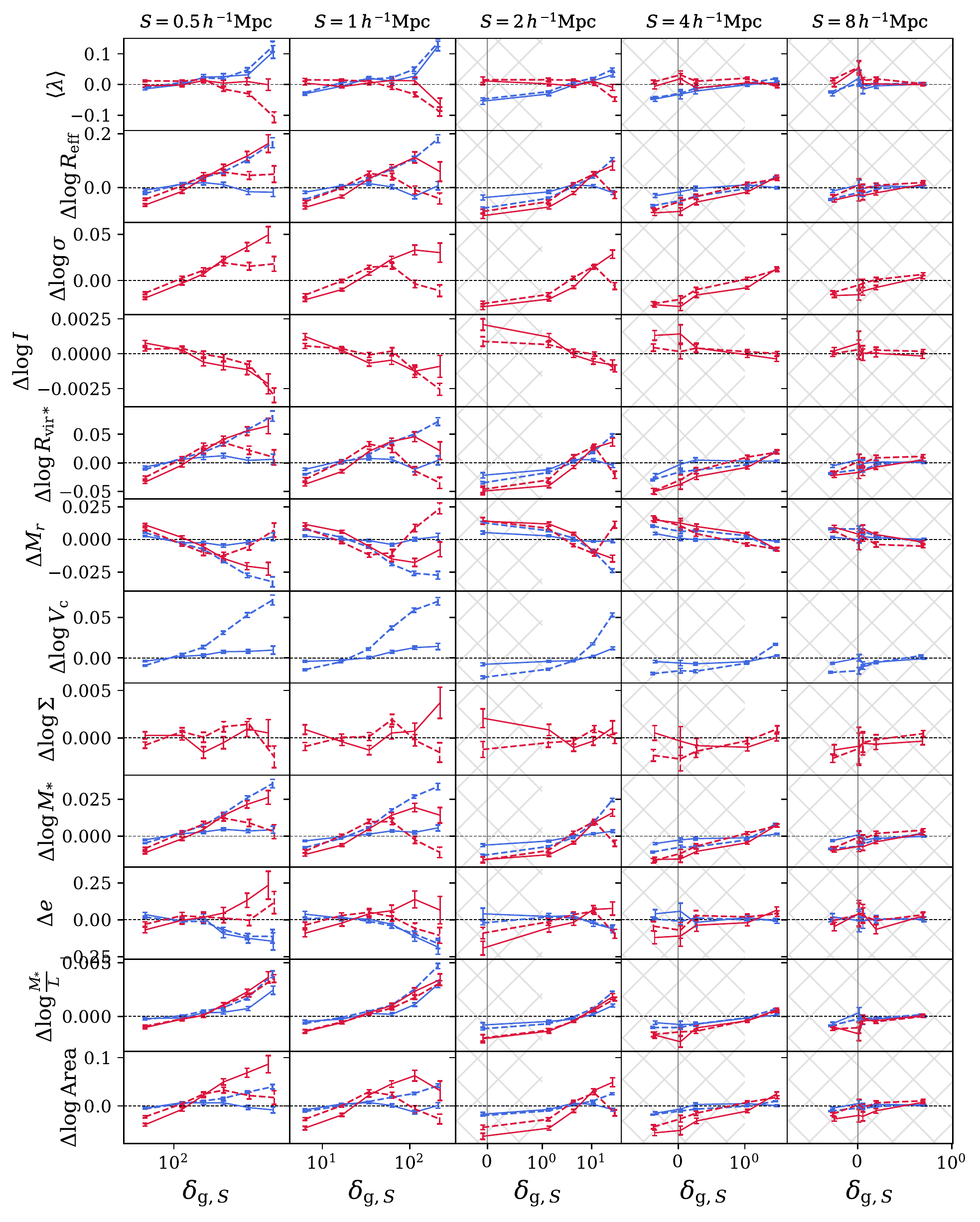}
    \caption{Correlations between galaxy sample (Table \ref{tab:sample_details}) average properties (\emph{rows}) and the local galaxy density contrast $\dgr$ within spheres of radius $S$ (\emph{columns}) -- thus columns increase the smoothing scale for the local density estimate, from left to right. Solid (dashed) curves give Masscut (Full) Fundamnetal Plane sample objects, with spiral (elliptical) samples shown in blue (red) -- see Table \ref{tab:sample_details} for sample details. Excepting the top row, which gives the density-binned mean FP residual $\meanl$, points give $\Delta{}X=\langle{}X\rangle_{\dgr} / |\langle{}X\rangle| - 1$, the fractional deviation of the mean of quantity $X$ in a local density bin from the mean of $X$ over the entire type/mass-selected sub-sample. Errors give the standard error on the mean. Hatching indicates linear axis spacing in the range $-1<\dgr<1$, with logarithmic spacing outside. Rows are, respectively, the Fundamental Plane radius residual $\lambda$, the measured log-effective radius, the velocity dispersion $\sigma$ for ellipticals, the surface brightness $I$ for ellipticals, the absolute magnitude, the circular velocity for spirals, the surface mass density $\Sigma$ for ellipticals, the stellar mass, the ellipticity $e$, the stellar mass-to-light ratio $M_{*}/L$, and the surface area (see Sect. \ref{sec:simulation} for property definitions). Elliptical and spiral galaxy samples exhibit distinct correlations of their properties with the local density, which are sensitive to the inclusion of lower-mass objects, and to the smoothing scale $S$ applied to the density field.
    }
    \label{fig:local_density_correlations}
\end{figure*}

\begin{figure*}
    \centering
    \includegraphics[width=\textwidth]{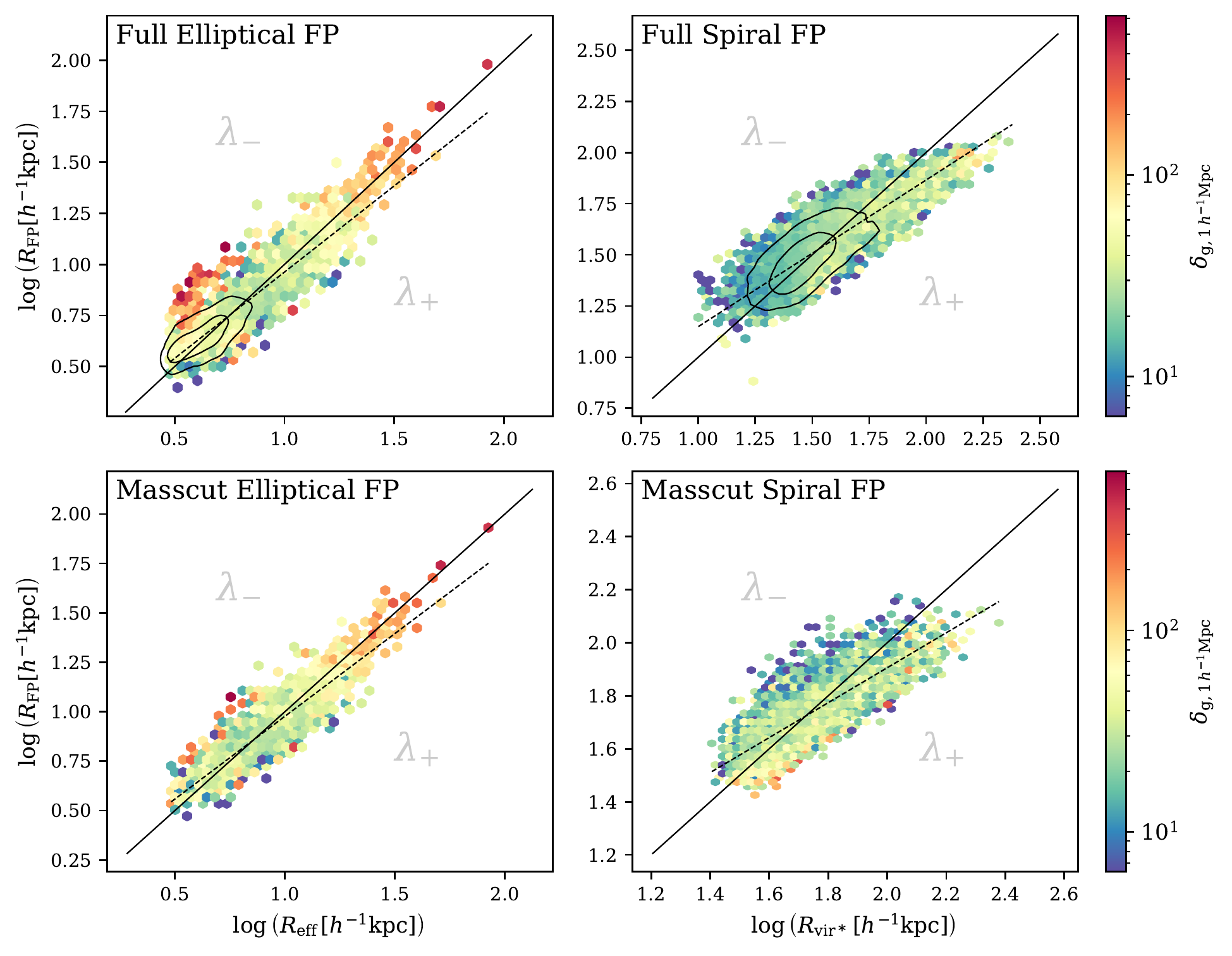}
    \caption{Elliptical (\emph{left}) and spiral (\emph{right}) Fundamental Planes, as in Fig. \ref{fig:fittedFPs}, but coloured here by the galaxy number density contrast $\dgr$ estimated for galaxies in spheres of radius $S=1\mpch$ (Sect. \ref{sec:environmental_dependence}). The $1:1$ relation, the linear fits to the points on each plane, and the contours showing the lower-mass objects' distributions are the same as those in Fig. \ref{fig:fittedFPs}, now given in black. Annotations $\lambda_{-}$ and $\lambda_{+}$ illustrate the sides of the plane corresponding to intrinsically small and large galaxies, respectively. Clear and opposite gradients in local density are seen across the Fundamental Planes of spiral and elliptical galaxies, with or without the application of the conservative mass selection (\emph{bottom}; see Sect. \ref{sec:simulation}).
    }
    \label{fig:FPs_by_density}
\end{figure*}

Fig. \ref{fig:ISCs} displays the three-dimensional two-point correlations measured for our various samples (Table \ref{tab:sample_details}) defined in the Horizon-AGN simulation box. The top (bottom) row shows correlations measured for spiral (elliptical) FP samples, and columns show the galaxy clustering $\xigg$, the $\lambda_+$ vs. $\lambda_-$ clustering difference $\dgg$, the lens density-FP sample size correlations $\xigl$, and the size auto-correlation $\xill$, respectively. Large points and solid lines depict conservative estimates, measured for FP and lens samples with a mass-selection $M_{*}>10^{10.5}\msol$ (Sect. \ref{sec:simulation}; \citealt{Hatfield2019}), whilst small points and dotted lines include all FP galaxies, and elliptical lenses down to $10^{10}\msol$. Grey shading marks scales $>17\mpch$, beyond which uncertainties are likely to be underestimated by our jackknife procedure (Sect. \ref{sec:intrinsic_size_correlations}) -- we thus exclude these data-points from reports of statistical significance.

We observe strong non-linearity in the galaxy clustering correlations \citep[see also][]{Lee2021c}, such that we are unable to make good fits of linear bias models to any of the measured $\xigg$. As such, we do not follow \cite{joachimi} and \cite{Singh2020} in fitting jointly to the clustering $\xigg\propto{}b^{2}_{\rm{g}}$ and density-size cross-correlation $\xigl\propto{}b_{\rm{g}}B$, thereby breaking a degeneracy between the galaxy bias $b_{\rm{g}}$ and the ISC parameter $B$; we prefer instead to model the $\lala$ correlations (see Sect. \ref{sec:magnification_magnification_results}).

\subsubsection{Spiral intrinsic size correlations}
\label{sec:spiral_ISCs}

Our measured spiral galaxy clustering reveals differences (Fig. \ref{fig:ISCs}; top-left panel) when the samples are split into intrinsically large $\lambda_+$ and small $\lambda_-$ objects for the measurement of the $\dgg$ function (top-middle-left panel). In the Masscut case, low signal-to-noise and inter-scale correlations result in a $\dgg$ signal that is not significantly non-zero ($0.6\sigma$; large points). However, in the Full FP case (small points) a $\dgg$ signal of similar form is detected at a significance of $5.9\sigma$, suggesting that, when lower-mass spirals ($M_{*}<10^{10.5}\msol$) are included, spiral galaxies with intrinsically large virial radii $\lambda_+$ are more strongly clustered than their small-radius counterparts $\lambda_-$ over scales $\lesssim10\mpch$.

This detection of a difference in clustering across the spiral FP is supported by a $5.2\sigma$ detection of a positive density-intrinsic size correlation $\xigl$ (top-middle-right panel), the significance of which is similarly reduced (to $1.6\sigma$) upon removal of the lower-mass spirals for the Masscut FP; thus larger spirals are more apt to be found at the peaks of the density distribution, as traced by the lens sample, with the significant detection again relying upon the inclusion of lower-mass spiral galaxies.

\cite{Cebrian2014} found $\sim10\%$ larger $\reff$ for low-$z$ late-type galaxies in the field, and \cite{Matharu2019} found $z\sim1$ star-forming cluster galaxies to be $17\%$ smaller than field spirals, both at fixed stellar mass. We consider the stellar particle virial radius $\rvir$ here, and find $\reff\sim{}(0.2-0.3)\,\rvir^{0.5-0.6}$ for spiral galaxies (scatter $\sim0.2$ dex; see Sect. \ref{sec:spiral_properties}), such that the variation of absolute radius with density is unlikely to differ between virial/effective radii. Thus, unless the \emph{intrinsic} variations in $\rvir$ and $\reff$ with density (akin to those found by fixing stellar mass; Fig. \ref{fig:FPs_by_stellar_mass}) have opposite signs, the simulation disagrees with the findings of \cite{Cebrian2014} and \cite{Matharu2019} -- modulo differences in sample selection, and in the different controlling variables ($M_{*}$ vs. expected radius). 

We make a $9\sigma$ detection of an intrinsic size auto-correlation $\xill$ (Fig. \ref{fig:ISCs}, top-right panel) for the Full spiral FP, and in contrast to $\dgg$ and $\xigl$, this signal remains significant at $\sim3.4\sigma$ even after removing lower-mass spirals. Thus the simulation strongly suggests that intrinsically large or small spiral galaxies tend to have similarly-sized neighbours, as reckoned by the stellar particle virial radius (Sect. \ref{sec:spiral_properties}).

As Fig. \ref{fig:lambda_vs_gxy_props} shows, any correlation between spiral $\lambda$ and stellar mass is weaker for the Full than for the Masscut spiral FP. It therefore seems unlikely that these trends are simple manifestations of mass-selection and changing galaxy bias, since the Full FP sample is far more evenly distributed on the $M_{*}-\lambda$ plane. The tilt in the spiral FP might offer some explanation, as it results in a tendency of intrinsically large spirals to also be absolutely large, and absolute size does correlate with stellar mass (the likely source of stellar mass-$\lambda$ correlations in Fig. \ref{fig:lambda_vs_gxy_props}). However, the galaxies responsible for these significant signals have typically small-to-intermediate virial radii in absolute terms, and are distributed far more symmetrically about the spiral Fundamental Plane (see Fig. \ref{fig:fittedFPs}; blue contours in top-right panel), such that the tilt alone cannot satisfactorily explain the array of significant signals from the Full spiral FP.

We thus conclude that spiral ISCs are in fact present in the simulation. Here we must reiterate that the lower-mass objects ($M_{*}<10^{10.5}\msol$; Sect. \ref{sec:simulation}) most strongly revealing these size correlations are contributors to a general underestimation of the galaxy clustering amplitude by Horizon-AGN, as compared with observations \citep{Hatfield2019}. However, their symmetric distribution around the FP, the lack of $\lambda$ vs. property correlations, and the persistence of the $\lala$ signal after limiting to higher-mass spirals, prevent us from discounting these correlations.

\subsubsection{Elliptical intrinsic size correlations}
\label{sec:elliptical_ISCs}

The signals we measure for elliptical FP samples are more sensitive to our analysis choices with respect to sample selections and the Fundamental Plane. We find that the usage of galaxy surface mass density $\Sigma$ in the FP \citep[as in][]{Rosito2020}, as opposed to the surface brightness $I$, effectively erases measurable ISCs when holding the galaxy sample constant. We also see great variability of FP fits and measured ISCs with respect to the inclusion of smaller, lower-mass elliptical galaxies in the simulation.

Briefly, relaxation of the global elliptical mass-selection $M_{*}>10^{10}\msol$ rapidly drives steep, positive $\dgg$ correlations on small scales, becoming negative at intermediate scales, whilst simultaneously erasing any $\xigl$ and $\xill$. Moreover, the total clustering signal $\xigg$ outstrips both the $\lambda_+$ and $\lambda_-$ clustering profiles on small scales, with $\lambda_+$ rising to dominate on large scales. Given the confused correlation picture which aligns poorly with literature findings; poorer FP fits and induced asymmetry in $\lambda$ (Sect. \ref{sec:additional_selections}); the steeply increasing stellar mass-function (as $M_{*}$ decreases from $10^{10}\msol$) of simulated ellipticals (Fig. \ref{fig:distributions}), which are known to be over-produced at the low-mass end \citep{Kaviraj2017}; and the aforementioned concerns around low-mass clustering in Horizon-AGN, we opt to exclude these objects from our Full FP sample (and Lens samples; Table \ref{tab:sample_details}), imposing that elliptical stellar masses $M_{*}>10^{10}\msol$.

For our fiducial choices, and upon inclusion of lower-mass ellipticals with $10^{10}<M_{*}/\msol<10^{10.5}$, we are in agreement with the observational findings of \cite{joachimi}. That work studied fainter elliptical galaxies at low redshifts ($z<0.2$), and made significant detections of negative $\dgg$ and ${\rm{g}\lambda}$, as well as a tentative detection of positive $\lala$. We reproduce the negative $\dgg$ detection at $\sim5.6\sigma$ when considering the Full FP sample, and this is accompanied by weak hints of negative $\xigl$ ($1.7\sigma$) and positive $\xill$ ($1.4\sigma$), wherein scales are highly correlated. Thus we observe ellipticals with intrinsically small effective radii (Sect. \ref{sec:elliptical_properties}) to be more clustered than their large counterparts, in opposition to the virial radius trend for spirals, and over a slightly broader range of scales. However, none of these signals are detected at $>0.7\sigma$ when the mass-selection $M_{*}>10^{10.5}\msol$ is imposed.

\cite{Singh2020} detected positive ${\rm{g}\lambda}$ correlations in higher-redshift ($0.16<z<0.7$) luminous red galaxy (LRG) data. However, their findings are compatible with ours and those of \cite{joachimi}, given that their sample is comparatively dominated by brighter, more massive objects; they were able to reproduce the negative signal seen by \cite{joachimi} by making selections in galaxy luminosity and colour. Interestingly, our Masscut FP signals weaken towards zero as compared with the Full FP signals in each case. This is agreeable with the findings of \cite{Singh2020}, since many of their low-luminosity selections -- bringing their sample more into parity with our own -- yield signals consistent with zero, or weakly negative ($\sim2\sigma$) in the faintest cases.

Both \cite{joachimi} and \cite{Singh2020} argued that the consistent assignment of central and satellite galaxies to $\lambda_+$ and $\lambda_-$, respectively, drove the ${\rm{g}\lambda}$ trend; that is, elliptical FPs and their residuals are dependent upon galaxies' environments \citep{Bernardi2003,DOnofrio2008,LaBarbera2010,Magoulas2012,Cappellari2013a,Hou2015,Singh2020,Howlett2022}. However, \cite{Saglia2010}, \cite{Saulder2019} and \cite{Singh2020} reported correlations between elliptical FP residuals and stellar mass/luminosity, which could partially account for a perceived environmental dependence.

In Horizon-AGN, we see no strong correlations between elliptical $\lambda$ and luminosity in Fig. \ref{fig:lambda_vs_gxy_props}. Further splitting of the elliptical galaxy samples in search of high-mass/luminosity signals is complicated by the already-small sample size -- we are unable to successfully fit an FP to more than $\sim6200$ ellipticals (Table \ref{tab:sample_details}, Sect. \ref{sec:fundamental_planes_results}). We thus explore measurable correlations between the properties of both elliptical and spiral galaxies in our FP samples with estimates of the local galaxy density contrast.

\subsection{Environmental dependence}
\label{sec:environmental_dependence}

We investigate the correlations of properties of our selected samples with their environments, using the local galaxy density contrast $\dgr$, in spheres of radius $S\mpch$, as a proxy. We note that the impact of tides imposed by the anistropic cosmic web structure will not be well-characterised by this isotropic density metric. Future analyses of simulated ISCs should thus consider environmental geometry and top-down scale coupling in galaxy formation and evolution, which has been shown to influence galaxy properties that are related to sizes \citep{Pichon2011,Codis2012,Codis2015,Laigle2015,Kraljic2018,Musso2018}.

We display property vs. $\dgr$ correlations in Fig. \ref{fig:local_density_correlations}, with blue (red) curves denoting spiral (elliptical) samples, and solid (dashed) curves denoting the Masscut (Full) FPs. Galaxies are geometrically-binned according to the number density (computed using all galaxies from Table \ref{tab:sample_details}) of spheres centred on their locations, and then normalised by the mean density in the box to give the galaxy density contrast $\dgr$. Thus from left to right in each panel, curves report statistics for low- to high-density regions, and the columns increase the smoothing scale from left to right. Points then give the fractional deviation of the density-binned mean property, relative to the un-binned mean property, with error bars equal to the appropriately propagated standard error on the mean. The $x$-axes are log-spaced, except where hatching indicates linear spacing between $-1 < \dgr < 1$.

The observed relations between $\meanl$ and $\dgr$ (Fig. \ref{fig:local_density_correlations}, top row) are consistent with our measured ISCs; high-density regions host spirals (ellipticals) which tend to have positive (negative) $\lambda$, hence we see opposite signs for spiral and elliptical $\dgg,\xigl$ in Fig. \ref{fig:ISCs}. The deprecation of signals on application of the mass-selection is also explained, as we see that the inclusion of lower-mass objects creates or steepens density-$\meanl$ correlations in most panels (dashed curves). We start by discussing elliptical galaxies, which are more sensitive to the mass-selection at the level of ISCs.

\subsubsection{Elliptical environmental dependence}
\label{sec:elliptical_environmental_dependence}

For elliptical galaxies, correlations are pronounced between $\dgr$ and effective radius $\reff$, velocity dispersion $\sigma$, virial radius $\rvir$, absolute magnitude $M_r$, stellar mass $M_{*}$, and to slightly lesser extents, surface brightness $I$ and ellipticity $e$, suggesting systematic trends in the early-type galaxy population between variably-dense environments in the simulation.

The qualitative forms of $\dgr-\{\reff,\sigma,I,M_r\}$ are each in agreement with the two-point correlations measured by \cite{Singh2020}. We note that the surface brightness $I$ correlates negatively with the local density, whilst the surface stellar mass density relation $\dgr-\Sigma$ is noisier, displaying no clear correlation. This is suggestive of variable elliptical mass-to-light ratios across different environments \citep{LaBarbera2010,Cappellari2013a,Cappellari2013,Suess2019a,Suess2019}, and may be related to the erasure of ISCs measured on the $\Sigma$-FP. Indeed, we see positive correlations of the stellar mass-to-light ratio $M_{*}/L$ with local density, in agreement with \cite{LaBarbera2010} (though we note that our stellar masses and luminosities are computed for all stellar particles assigned to a galaxy, and not at any principled radius; \citealt{Cappellari2013a}). Coupled with the positive $\dgr-M_{*}$ correlation, we infer that the surface area of our ellipticals generally increases with local density, and this is confirmed by the measured $\dgr-$surface area correlation (bottom row).

However, the relaxation of the mass-selection $M_{*}>10^{10.5}\msol$ in the elliptical FP sample can be seen in Fig. \ref{fig:local_density_correlations} to slash many of these trends (solid vs. dashed red curves). The additional lower-mass galaxies primarily influence the mean properties of high-density environments by dragging them back towards the global means (given by zero-lines in the figure), with the notable exceptions of $\meanl$, for which a negative $\meanl$ at high-$\dgr$ is induced by the lower-mass objects, and $M_{*}/L$ and $I$, which remain consistent.

Thus, relative to the Masscut sample, regions of high density in the Full FP sample are seen to contain more compact, fainter, lower stellar mass, lower velocity dispersion elliptical galaxies \citep[as seen by][]{Poggianti2013,Cappellari2013a,Baldry2020}. These changes in density-mean property gradients could be sourced by the addition of e.g. larger ellipticals to intermediate-density environments, but given that the lower-mass galaxies are known to be of lower absolute radius (Fig. \ref{fig:distributions}), we can infer that they are indeed preferentially located in high-density environments. The expanded population of ellipticals in dense environments then induces the gradient in $\dgr-\meanl$, which manifests as non-zero, negative $\dgg$ and $\xigl$ signals for the Full FP sample, where the former is statistically significant.

\subsubsection{Spiral environment dependence}
\label{sec:spiral_environmental_dependence}

Our Masscut spiral FP samples show positive correlations between the local density and the circular velocity $V_{\rm{c}}$, the stellar mass $M_{*}$, and the stellar mass-to-light ratio $M_{*}/L$, which greatly strengthen upon relaxation of the mass-selection (dashed vs. solid blue lines). Correlations between $\dgr$ and effective/virial radii (absolute magnitudes) meanwhile are weak for the Masscut sample, and strongly positive (negative) for the Full sample, suggesting that the Masscut sample is more homogenised, in terms of absolute size, across density regimes.

The lack of environmental variation in the radii of high-mass spirals $M_{*}>10^{10.5}\msol$ is in agreement with the observational findings of \cite{Maltby2010} and \cite{Lani2013}, who saw no strong evidence for any such scaling. However, \cite{Maltby2010}, \cite{Cebrian2014} and \cite{Matharu2019} also saw in observations that lower-mass spirals tended to be larger in the field than in dense environments, which may disagree with what we see when relaxing the mass-selection in our simulated samples (dashed blue lines). Whilst the correlation of absolute size with density may be explained by the matching stellar mass-density correlation, a weaker $\meanl-\dgr$ correlation persists, suggesting that (modulo $\meanl-R_{\rm eff/vir*}$ correlations; Fig. \ref{fig:lambda_vs_gxy_props}) spirals are larger in high-density environments at fixed stellar mass (Fig. \ref{fig:FPs_by_stellar_mass}). 
Trends showing increased luminosity at higher densities for the Full sample are agreeable with some studies of the environmental dependence of the late-type luminosity function \citep{Mo2004,Croton2005,Zucca2009,Eardley2015}, whilst the lack of correlation seen for the Masscut sample agrees with others \citep{Tempel2011,Zandivarez2011,Poudel2016}. More detailed investigations of these differences are beyond the scope of this work.

Noting that the radii of lower-mass spirals are smaller than those existing only in the Masscut sample (Fig. \ref{fig:distributions}; low-mass also have also lower $V_{\rm{c}}$, not shown), the changing density-radius trends must be driven by the addition of small-radius objects to the field. This offers some explanation for the loss of signal significance upon application of the mass-selection, seen in Sect. \ref{sec:spiral_ISCs}; whilst the $\dgr-\meanl$ trend remains for the Masscut spiral sample, the measured correlations $\dgg$ and $\xigl$ are rendered statistically insignificant by the loss of signal-to-noise due to the removal of $\sim75\%$ (Table \ref{tab:sample_details}) of the source objects, preferentially from the low-density field. Meanwhile the loss of signal-to-noise is less severe for the $\xill$ correlation, which relies less upon a broad sampling of environments, and is thus maintained at $>3\sigma$ significance through the mass-selection (though there are some indications that the Masscut spiral $\xill$ may be contaminated by $\lambda$-property correlations; see Appendix \ref{sec:extended_fundamental_planes}).

\subsection{Intrinsic size correlations summarised}
\label{sec:ISCs_summarised}

We further elucidate the intrinsic/absolute size-density trends by considering the FPs again, now coloured by the local density estimates from Sect. \ref{sec:environmental_dependence}, shown in Fig. \ref{fig:FPs_by_density}. For the smoothing scale $S=1\mpch$, one clearly sees gradients in $\dgr$ in each panel, in directions almost orthogonal to the $1:1$ relation, here given in black, with $\lambda_{-},\lambda_{+}$ annotations to denote the intrinsically small and large radii. The dashed black lines and contours illustrate the tilts of the planes, and the distribution of lower-mass objects removed from the Masscut FPs, respectively (as described in Sect. \ref{sec:fundamental_planes_results}).

The tendency of absolutely small, intrinsically small (large) ellipticals to inhabit more (less) dense environments is seen in both of the left-panels, and is more pronounced when the smoothing scale is reduced to $S=0.5\mpch$ but quickly weakens for $S\geq2\mpch$ (not shown) -- signifying that more localised structures are the drivers of this environmental dependence.

Similarly, one begins to see hints of an opposite gradient in the top-right panel, for Full FP spirals, and even more so in the Masscut case (bottom-right). For spirals, the gradients are weaker for $S=0.5\mpch$, but stronger for $S=2,4\mpch$, and still visible for $S=8\mpch$; the structures driving this correlation of local density with FP residuals are thus more extended than those driving the elliptical correlations.

The evolving galaxy size-mass-morphology distribution is likely to have a complex dependence upon the anisotropic cosmic web environment of voids, walls, filaments, saddle points, and nodes, where the relative efficiency and vorticity of gas flows dictate the build-up of galactic angular momenta \citep{Pichon2011,Codis2012,Welker2014,Codis2015,Laigle2015}; the bulk velocity flows transport galaxies through the different environments over cosmic time \citep{Codis2015,Kraljic2018,Laigle2018}; and the variable efficiency of mass accretion and mergers strongly influence galaxies' formation and evolution \citep{Welker2017,Musso2018}. Each such process has the potential to force galaxies out of kinetic equilibrium, and/or influence the new equilibrium to be reached after relaxation; indeed the intrinsic size distribution may be dependent upon the balance of different kinetic equilibria and dis-equilibria.

We defer a detailed analysis of the intrinsic size distribution across anisotropic cosmic environments to future work. For now we turn to \cite{Welker2017}, who studied the impacts of different stellar growth processes upon galaxy size and morphological evolution over $z=1-5$ in Horizon-AGN. Whilst we are extrapolating their findings from $z=1-5$ down to $z=0$ for our interpretations here, \cite{Welker2017} found mergers of any kind to be rare, violent events, occurring on average only twice in the history of a galaxy over the epoch $z=1-5$. Given that the merger rate decreases with cosmic time \citep{Rodriguez-Gomez2015}, we assume that their conclusions are largely applicable to our work here.

They found that multiple minor, gas-deprived mergers had similar effects to major mergers in terms of destroying disc structures and forming spheroids. Minor, gas-rich mergers were even seen to cause contraction of the remnant objects \citep[compatibly with the gas compaction paradigm of][]{Dekel2014,Inoue2016}. Thus wet mergers are a possible pathway for the compact elliptical galaxies driving our measured correlations, provided that they are not consumed by later mergers. One does see that the absolutely-and-intrinsically small galaxies constitute a minority of the supplementary objects in the Full elliptical FP, as they sit largely outside of the black contours in Fig. \ref{fig:FPs_by_density} (top-right panel); these objects are indeed rare at low redshift, possibly having been consumed by central galaxies over cosmic time \citep{Matharu2019,Baldry2020}.

Meanwhile, \cite{Welker2017} also found that cold gas flows and consequent \insitu star formation, as well as minor mergers, tended to flatten lower-mass $M_{*}<10^{10.5}\msol$ spheroidal galaxies along their minor axes, with the former potentially prompting the (re-)formation of discs among galaxies that avoid subsequent mergers. Regardless of further mergers or disc reformation, if these processes sufficiently tip the balance of $V/\sigma$, we might then classify these objects as spirals. At fixed stellar mass, spheroidal galaxies are larger in the simulation \citep{Dubois2016}\footnote{We note that higher-resolution hydro-simulations produce more compact $z=2$ galaxies at fixed stellar mass, owing to improved modelling of gas flows \citep{Chabanier2020}, in better agreement with observations that show discs to be larger than spheroids at fixed stellar mass \citep{VanDerWel2014,Kawinwanichakij2021,Nedkova2021}.}, and expected to exist closer to filaments/nodes \citep{Kraljic2018,Laigle2018}, than typical discs at the same stellar mass; disc reformation is therefore a possible pathway for the intrinsically large spirals sourcing our ISCs.

Future efforts to dis/confirm our interpretation of ISCs in hydrodynamical cosmological simulations could focus upon the merger histories of simulated objects, as a function of the anisotropic cosmic web environment, extending to higher-redshift snapshots to assess the interplay of mergers and morphological transitions, and the associated change in profile of ISCs. Indeed, one might seek to more finely bin galaxy samples using a metric that directly discriminates between stages of evolution, such as $V/\sigma$, age \citep[see also][]{Lu2020}, or metallicity \citep{LaBarbera2010}, or else according to galaxies' local, anisotropic environments. Our work here should motivate a search for smooth trends in ISCs with respect to such variables, as objects evolve through the stages, and migrate through the cosmic web environments, that we have coarsely probed here.

The recently-completed Horizon Run 5 \citep{Lee2021c} should be especially useful for answering remaining questions with regards to galaxy intrinsic size correlations in hydrodynamical cosmological simulations, with a cuboid box having $10\times$ the volume explored here, a significant number of galaxy clusters, and improvements to sub-grid physics implementation that could dispel some of our limiting concerns with regards to the behaviour of low-mass galaxies.

Before moving on, we note here some limitations of our analysis. Our usage of stellar particle virial radii as size metrics for spiral galaxies is non-optimal, as these may not correspond well to common observational spiral radii -- such as half-mass/light radii, or disc scale-lengths -- and would be far more difficult to obtain for real data. Moreover, the correlations between FP residuals $\lambda$ and spiral galaxy properties such as effective/virial radii are causes for concern. Moving from the Masscut to the Full FP sample generally increases the significance of ISCs without dramatic changes to their form (Fig. \ref{fig:ISCs}), whilst also reducing the $\lambda$-property correlations (Fig. \ref{fig:lambda_vs_gxy_props}). However, the clustering statistics of these lower-mass objects are questionable \citep{Hatfield2019}, and this should also temper our assessment of the elliptical ISCs, where the significant Full FP $\dgg$ signal is heavily sourced by these galaxies.

We therefore explore some extended FPs and ISCs in Appendix \ref{sec:extended_fundamental_planes}, where additional FP variables serve to reduce FP scatter/tilting and $\lambda$-property correlations, for both spiral and elliptical samples. Whilst some of the resulting ISC signals lose statistical significance relative to our fiducial analysis, others gain greatly, and none are so different in form as to be of concern. Our conclusions with respect to the variability of intrinsic sizes with environment and morphology are maintained, and we refer the reader to the appendix for more details.

\subsection{Density-magnification contamination}
\label{sec:density_magnification_results}

\begin{figure*}
    \centering
    \includegraphics[width=0.85\textwidth]{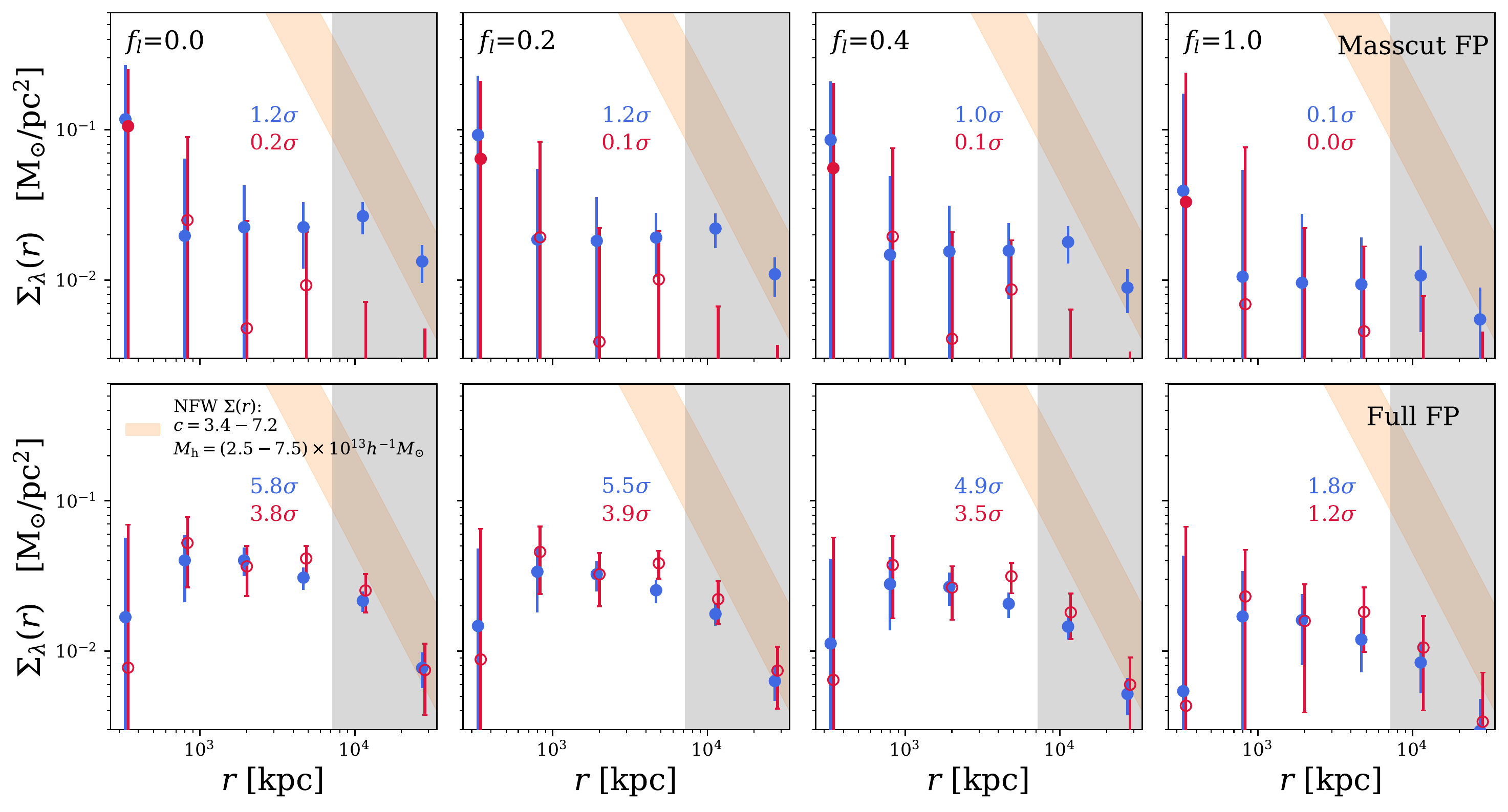}
    \caption{Predicted intrinsic contributions $\Sigma_{\lambda}(r)$ to the projected surface density $\Sigma(r)$, as measured from angular galaxy position-size correlations in the presence of catastrophic photometric redshift outliers (see Sect. \ref{sec:density_magnification_contamination}). Contributions from spiral (elliptical)  intrinsic sizes are shown in blue (red). Columns give different assumed values for the fraction $f_{l}$ of `source' galaxies that are in reality co-located with $z\sim 0.2$ lenses, whilst rows give the systematic contributor signal estimated for more conservative Masscut Fundamental Plane samples (\emph{top}; Sect. \ref{sec:simulation}), and for Full Fundamental Plane samples (\emph{bottom}). Data points and error bars correspond to the mean signals and standard deviations over 1000 bootstrap samples of the size catalogues. Coloured numbers give the significance of these estimates, computed with exclusion of the grey-shaded region ($\gtrsim 5\mpch$), and neglecting cosmic variance. 
    The orange shaded curve illustrates a range of NFW halo profile \citep*{NFW96,NFW97} expectations for $\Sigma(r)$, corresponding to $\pm1\sigma$ constraints on luminous red galaxy halo masses $M_{\rm h}$ and concentrations $c$ from \citet{Mandelbaum2006a}. The predictions for contaminants (data points) are seen to be of comparable amplitudes to the data-driven NFW model (orange curve) for $r\gtrsim 10\mpc$, particularly when the interloper fraction $f_l$ is underestimated (for this mock measurement, the true $f_l$ is exactly unity).
    }
    \label{fig:Sigma_lambda}
\end{figure*}

We now turn to our mock estimation of the intrinsic contamination affecting a lensing magnification measurement of the projected surface density $\Sigma(r)$ (detailed in Sect. \ref{sec:density_magnification_contamination}). Fig. \ref{fig:Sigma_lambda} displays our predictions of the systematic contribution $\Sigma_\lambda(r)$, as estimated via galaxy size-based cosmic convergence statistics \citep{Huff2011}, given different assumed values for the photo-$z$ interloper fraction $f_l$ (columns). Blue (red) data-points give the estimated contributions from spiral (elliptical) galaxies, with the top row featuring our Masscut FPs, and with Full FP signals on the bottom. Open points denote negative values, and the grey shading marks scales $>5\mpch$, beyond which our neglect of periodic boundary conditions renders predictions untrustworthy -- quoted significances (in-panel coloured numbers, without cosmic variance) thus exclude these scales. A range of NFW profile \citep*{NFW96,NFW97} predictions for $\Sigma(r)$, corresponding to $1\sigma$ confidence intervals constrained by \cite{Mandelbaum2006a} are illustrated as orange shading -- these are similar to the model depicted by \cite{Huff2011}, which was seen to accurately predict their measured $\Sigma(r)$ signal.

In the case of Masscut FP residuals, our predictions are consistent with null signals in all cases, as expected given the lack of significant ${\rm{g}}\lambda$ signal detections within the simulation box (Fig. \ref{fig:ISCs}; large points). However, when relaxing the mass-selection, those intrinsic size-density contributions become significantly non-zero, with their strength modulated by the assumed value of $f_l$. For our estimation procedure, the true interloper fraction is unity; all galaxies entering the estimator (Eq. \ref{eq:Sigma_lambda_estimator}) are coincident in redshift. In the bottom-right panel, one sees that the correct value for $f_l$ can mitigate most of the systematic signal, lowering the significance from $>5.5\sigma$ ($\sim3.8\sigma$) from spirals (ellipticals), when $f_l$ is severely underestimated, to $\sim1.8\sigma$ ($1.2\sigma$).

The correction works as follows: a population of interloper objects from redshift $z_l$ is thought to be at $z_s$, and its sizes thought to trace the convergence field. The intrinsic sizes of the population are correlated with the large-scale structure at $z_l$, and thus $\Delta\log{}R$ (Eqs. \ref{eq:huffgraves_kappa} \& \ref{eq:estimator_lambdahat}) correlates with lens sample positions despite the lack of convergence-derived galaxy size fluctuations. The $\Delta\log{}R_{\rm{err}}$ term (Eq. \ref{eq:FP_error_term}) is positive, because object redshifts have been overestimated\footnote{We note that underestimated redshifts could cause a negative correction term, which one would expect to hamper any mitigation of ISCs.} $z_s>z_l$ and we are far below the redshifts where angular diameter distances $d_{\rm{a}}(z)$ turn over; thus $f_l$ weights a term that serves to make $\Delta\log{}R$ smaller, or more negative. When raised to an exponent to yield the estimator for $\hat\lambda$ ($\equiv\widehat{\kappa+\lambda}$ in a real analysis; Eq. \ref{eq:estimator_wISC}), $f_l\rightarrow1$ maximises a suppression of the systematic contribution; that is, the photo-$z$ correction term mitigates ISCs by chance.

However, this correction involves discarding large amounts of data from small-angle regimes with the highest signal-to-noise for the convergence. It also cannot fully disentangle $\lambda$ from $\kappa$ in the presence of ISCs, as Eq. \ref{eq:estimator_wISC} cannot differentiate between convergence and intrinsic size variation. If the spatial average of $\Delta\log{}R$ is non-zero due to intrinsic correlations with the `foreground' structure, then depending upon the strength of the underlying intrinsic correlation, and the proportion of intrinsically correlated galaxies under consideration, even a correct determination of $f_l$ may not completely suppress $\Sigma_\lambda$, as can be tentatively seen in the bottom-right panel of Fig. \ref{fig:Sigma_lambda}. It would be preferable for a magnification probe analysis to forward-model both the ISC and photo-$z$ effects, to allow for a cleaner extraction of the convergence $\kappa$.

\cite{Sheldon2004} found $\Delta\Sigma(r)$ to be  $<1\,h\,\msol{\parsec^{-2}}$ at $\sim10\mpch$, and $\Sigma(r)$ follows a steeper power law with respect to $r$. Thus smooth extensions of our Full FP $\Sigma_\lambda$ estimates to scales of order $10\mpc$ and beyond\footnote{The grey-shaded region in Fig. \ref{fig:Sigma_lambda} excludes scales $\gtrsim5\mpch$, beyond which our neglect of periodic boundaries (Sect. \ref{sec:density_magnification_contamination}) is likely to become important; the amplitudes of the final two points (the last especially) are thus likely to be underestimated.} would be likely to impinge upon the lensing signal, as illustrated by the orange shaded curve in Fig. \ref{fig:Sigma_lambda}.

Interestingly, the noisier elliptical size-density correlation is of a similar amplitude to that from spirals but with an opposite sign. Given the divergent intrinsic properties of spiral and elliptical galaxies, and the current lack of a unified FP to describe joint scaling relations for their radii, one might assume that they will be kept separate for a size-based cosmic convergence analysis \citep{Alsing2015}. Analysing both samples for magnification in parallel would also provide a beneficial cross-checking mechanism, where estimates of $\Sigma(r)$ should be consistent for the same lenses when traced by sources of either morphological type \citep[see e.g.][for a colour-split consistency analysis of cosmic shear]{Li2021}. Our work suggests that such an analysis will be contending with the fidelity of morphology-dependent ISC modelling for accurate determinations of $\Sigma(r)$ on large scales.

We recall here that this estimation of an intrinsic contamination to $\Sigma(r)$ derives from a single simulation snapshot at $z=0.06$, assuming no significant redshift evolution of $f_l$, or of the density-intrinsic size correlation out to $z\sim0.2$. These strong assumptions are convenient for this work, since we analyse only one simulation snapshot situated at $z=0.06$, whilst optimal lens samples are to be defined at higher redshifts, allowing for more efficient lensing of yet-higher redshift background sources. Future studies of ISC phenomena in simulations should extend the redshift baseline, consider deriving light-cones, and investigate functional form(s) for $f_l(z_l, z_s, \vartheta)$ motivated by such work as \cite{Stolzner2022}.

We note that an interesting route to study the impacts of ISCs in lensing magnification would be to apply the self-calibration approach of \citep{Zhang2010,Zhang2010a}. As shown for real data by \cite{Yao2020a,Yao2020}, the radial asymmetry of the gravitational lensing effect, combined with photometric redshift information, can be leveraged to extract statistically isotropic intrinsic alignment contributions from galaxy position-shear correlations in a model-independent way. This reasoning should also hold for intrinsic size correlations, and offers a promising alternative mode of study.

\subsection{Magnification-magnification contamination}
\label{sec:magnification_magnification_results}

\begin{figure}
    \centering
    \includegraphics[width=\columnwidth]{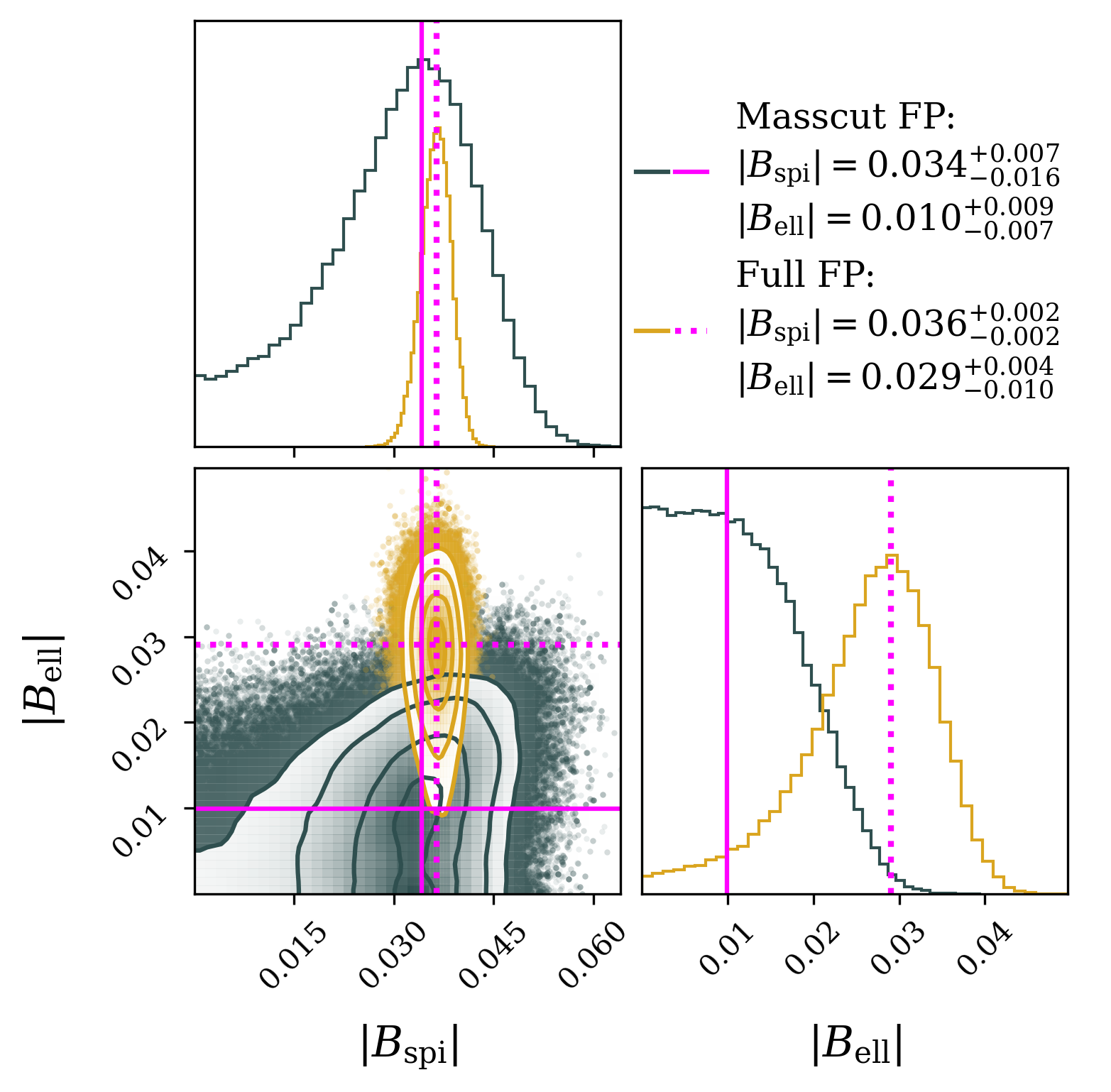}
    \caption{Constraints upon the linear, absolute intrinsic size parameters $|B_{\rm{spi/ell}}|$, as fitted (see Sect. \ref{sec:magnification_magnification_contamination}) to the auto-correlation function $\xill(r)=B^2\xi_{\rm{\delta\delta}}$ (Fig. \ref{fig:ISCs}) of Fundamental Plane radius residuals $\lambda$ for spiral and elliptical galaxies in the Horizon-AGN simulation box. The Full spiral Fundamental Plane model amplitude is constrained to be non-zero at high signifcance, whilst Full elliptical and Masscut Fundamental Plane constraints are each consistent with zero at $\leq3\sigma$.}
    \label{fig:ll_fits}
\end{figure}

\begin{figure*}
    \centering
    \includegraphics[width=\textwidth]{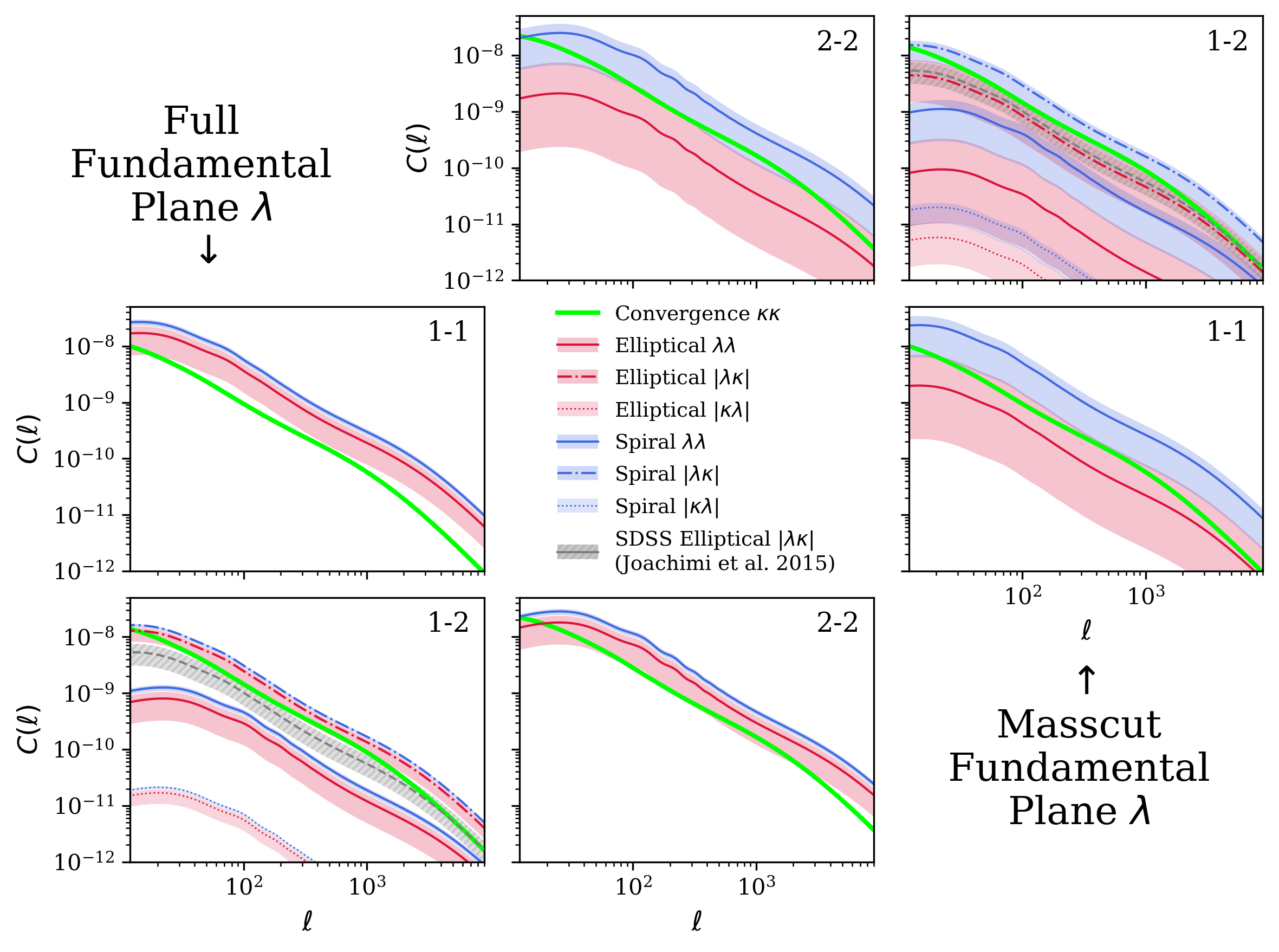}
    \caption{Predicted intrinsic contributions to size-based angular convergence power spectrum estimates $C(\ell)$ derived from fits of a phenomenological model (Sect. \ref{sec:magnification_magnification_contamination}) for intrinsic size correlations $\xill$ (Fig. \ref{fig:ISCs}). Shown in green are the cosmic convergence spectra $\kk$, whilst in blue (red) are the intrinsic, systematic contributions from spiral (elliptical) galaxies, with auto- and cross-correlations shown for two tomographic redshift samples at $z\sim 0.4,0.8$. Solid red/blue curves give the Fundamental Plane residual auto-correlations $\lala$; dot-dashed curves the absolute value of the intrinsic size-convergence cross-correlation $|\lk|$; dotted curves the reverse cross-correlation $|\kappa\lambda|$; and grey dashed curves the $|\lk|$ prediction from \citet{joachimi}. The envelope around each curve gives the $\sigma_{68}$ confidence interval corresponding to the model constraint ($\pm1\sigma$ for the \citealt{joachimi} prediction). Predictions from more conservative Masscut Fundamental Planes are shown in the upper triangle, whilst the Full Fundamental Plane results are given in the lower triangle. Contaminant contributions to the convergence power spectrum are seen to be comparable to the cosmological signal in several cases, and most significantly for the Full elliptical and spiral Fundamental Plane residuals.
    }
    \label{fig:kappakappa_masscut}
\end{figure*}

The results of fitting our phenomenological model to measured $\xill$ correlations (Sect. \ref{sec:magnification_magnification_contamination}) are shown in Fig. \ref{fig:ISCs} as pink solid and dotted curves, for the Masscut and Full FP samples, respectively. We note that this simple model seems to fit the measured Full spiral FP $\xill$ correlation remarkably well, extending far into the non-linear regime despite being limited scales $r>3\mpch$. Considering the less conservative sample selection, and without a physical motivation for the model, we are reluctant to make any strong conclusions regarding the quality of this particular fit, though the fit to the Masscut spiral $\xill$ is comfortably consistent. Conversely, the model prefers a significantly larger amplitude for the Full elliptical FP signal than for the Masscut signal, where the latter fit runs into the zero-prior.

Confidence contours depicting our phenomenological model constraints are shown in Fig. \ref{fig:ll_fits}, with grey denoting the Masscut FP fits, and gold the fits to Full FP signals. Maximum likelihood points (corresponding to the curves in Fig. \ref{fig:ISCs}) are indicated in each case by pink lines, with solid (dotted) lines giving the Masscut (Full) FP results. The reduced\footnote{Taking either one parameter and three data-points, thus two degrees of freedom per individual $\xill$, or else two parameters and six data-points, and thus four degrees of freedom for the joint computation.} $\chi^2$ is $0.33$ ($0.45$) for spiral (elliptical) $\xill$ from the Masscut FP, or $0.30$ when computed jointly thus accounting for the small covariance between individual $\lala$ signals. For the Full FP signals, the reduced $\chi^2$ is $0.37$ ($0.55$) for spiral (elliptical) $\xill$, or $0.49$ when computed jointly. We note that the inter- and intra-$\xill$ correlations are larger in the Full FP case, and feature an anti-correlation between spiral/elliptical signals. 

The results of fitting are negligibly different when performed jointly or separately. We report for the Masscut FP: $|B_{\rm{spi}}|=0.034^{+0.007}_{-0.016},\,|B_{\rm{ell}}|=0.010^{+0.009}_{-0.007}$, and for the Full FP: $|B_{\rm{spi}}|=0.036^{+0.002}_{-0.002},\,|B_{\rm{ell}}|=0.029^{+0.004}_{-0.010}$, where errors refer to the difference between the maximum likelihood points and the 16th, 84th percentiles (the $\sigma_{68}$ interval in the following figures) of Monte Carlo Markov Chains after an appropriate burn-in phase.

Thus our phenomenological model prefers a non-zero intrinsic size-matter density coupling $B_{\rm{spi}}$ at nearly $\gtrsim2\sigma$ in the Masscut case, and $\sim18\sigma$ for the Full FP, where the two are comfortably consistent. Conversely, the elliptical coupling $B_{\rm{ell}}$ runs into the prior $|B|\geq0$ in the Masscut case, and is nearly $3\sigma$ greater than zero for the Full FP. Each constitutes a good fit to the relevant signal, but we caution moderation, as this model is not physically motivated -- \cite{joachimi} used a very similar model and were unable to make good fits to their elliptical FP size residual auto-correlations $\lala$. Future work should revisit these signals with a more motivated description (e.g. the EFT model for scalars; \citealt{Vlah2019}, or the size-luminosity halo model of \citealt{Ciarlariello2015,Ciarlariello2016}).

As described in Sect. \ref{sec:magnification_magnification_contamination}, we translate these model constraints into predictions of angular power spectra (Eqs. \ref{eq:Cells}) describing the intrinsic size auto-correlation $\lala$, and intrinsic size-convergence cross-correlation $\lambda\kappa$, for two redshift samples at $z=0.4,0.8$ \citep{joachimi}, displaying the results in Fig. \ref{fig:kappakappa_masscut}. The figure shows systematic $C(\ell)$'s (red/blue) corresponding to the best-fit model parameters, with the $\sigma_{68}$ intervals given by shading, in comparison with the cosmic convergence power spectrum $C_{\kk}(\ell)$, shown in green. Results derived from Masscut (Full) FP residuals are given in the upper (lower) triangle.

The $\lala$ correlation (solid curves) must be non-negative, and we see that the predicted amplitude of $\lala$ signals from Masscut ellipticals is about an order of magnitude smaller than the convergence signal in the tomographic auto-correlations (middle- \& bottom-left, top- \& middle-right), though we remind the reader that $B_{\rm{ell}}$ from the Masscut FP runs into the prior $|B|\geq0$; these predictions are consistent with null signals.

The Masscut spiral $\lala$ signal, however, is likely to dominate over the convergence; this from the more conservative Masscut selection is potentially concerning for future studies of cosmic convergence. Since higher redshifts are dominated by late-type galaxies, most of the signal-to-noise in a weak lensing analysis is derived from these objects, and excluding them would make the cosmological signal much more difficult to detect -- perhaps especially in the case of size-based convergence, as the size dispersion is thought to be larger than the shear dispersion \citep{Huff2011}, placing even more stringent requirements upon the sample number density.

In the tomographic cross-correlation (bottom-left \& top-right), the $\lala$ contribution weakens substantially, owing to the small overlap between redshift distributions. The dominant systematic signal now comes from the intrinsic size-convergence cross-correlations $\lk$ (dot-dashed curves). Here, the sign of $B$ becomes important. We might assume (more securely for the Full FPs) that $B_{\rm{spi}}$ is positive, and $B_{\rm{ell}}$ negative, given the forms of $\dgg$ and $\xigl$ correlations seen in Fig. \ref{fig:ISCs}. We note, however, that our model strictly constrains only the absolute values of these parameters, and that each of the Masscut elliptical ISCs is consistent with a null detection. Thus the cross-correlation predictions are for absolute contributions $|\lk|$.

The spiral contribution $|\lk|$ remains likely to dominate over the cosmological signal. We display the ${\rm{g}\lambda}$-derived constraint $B_{\rm{ell}}=-0.012\pm{0.005}$ from \cite{joachimi} for comparison as a grey dashed curve with $1\sigma$ uncertainty given by shading/hatching. Our constraints upon $B_{\rm{ell}}$, and corresponding predictions for $|\lk|$, are consistent with that from \cite{joachimi} at 68\% confidence for the Masscut FP, whilst our Full FP predicts a stronger contamination. The small $|\kappa\lambda|$ contribution (dotted curves), from foreground convergence and background ISCs, derives from a very small overlap of corresponding lensing and number kernels (Eq. \ref{eq:Cells}; middle), and is unlikely to be of concern (similarly to the intrinsic alignment IG correlation in cosmic shear).

Each contamination prediction is strengthened when turning to the Full FP results, with spiral $\lala$ and $|\lk|$ signals factors of $2-3$ larger than the convergence at high significance. Meanwhile, elliptical $\lala$ and $|\lk|$ rise to become comparable to, or dominant over the convergence. 

%--------------------------------------------------------------------

\section{Conclusions}
\label{sec:conclusions}

We have explored the intrinsic size correlations (ISCs) of spiral and elliptical galaxies in the $z=0.06$ snapshot of the cosmological hydrodynamical simulation Horizon-AGN, as traced by Fundamental Plane (FP) radius residuals $\lambda$.

We defined morphological galaxy samples according to the ratio of stellar particles' average tangential velocity and velocity dispersion, with small (large) ratios tracing spiral (elliptical) populations. Considering the clustering deficits observed for low-mass Horizon-AGN objects by \cite{Hatfield2019}, we further divided the morphological samples into a conservative, higher-mass subset, `Masscut', with $M_{*}>10^{10.5}\msol$, and `Full' samples including the lower-mass galaxies (limited to $M_{*}>10^{10}\msol$ for ellipticals).

For elliptical galaxies we followed standard practice, with the effective radius $\reff$, velocity dispersion $\sigma$, and surface brightness $I$ forming the FP. We also explored FPs substituting the surface stellar mass density $\Sigma$ for the surface brightness $I$ \citep[as done by][for their fiducial planes]{Rosito2020}, and found that the radius residuals from this plane were not significantly correlated with the large-scale structure of the simulation, in disagreement with observational studies where such elliptical ISCs were measured at high significance  \citep{joachimi,Singh2020}. Unless otherwise stated, we henceforth refer to results derived from the surface brightness plane, which are far more agreeable with literature work.

We were inspired by \cite{Shen2002} to use a spiral galaxy radius estimate to tighten the scatter about the Tully-Fisher scaling of circular velocity $V_{\rm{c}}$ with absolute magnitude $M_{r}$, finding the virial radius of stellar particles $\rvir$ to yield planes with manageable degrees of tilting, and scatters comparable to those of the elliptical planes. Lower-mass spirals in the Full FP sample were seen to be particularly well-described, with the tilting of the FP featuring mainly at the high-mass end.

We defined the intrinsic galaxy size fluctuation $\lambda$ as the natural logarithm of the ratio of measured-to-predicted (by the fitted FP) radii for each object, finding these to be approximately normally distributed, and having similar distributions between spiral and elliptical, Full and Masscut FPs. The coefficients and root-mean-square deviations of our residuals $\lambda$ were seen to be comparable to those of elliptical FP residuals from the literature, measured for galaxies observed in reality and in simulations.

We measured two-point galaxy correlation functions as functions of three-dimensional separations $r$, weighted by the FP residuals $\lambda$: the density-intrinsic size correlation $\xigl(r)$, and the intrinsic size auto-correlation $\xill(r)$. We also measured the galaxy clustering correlation function $\xigg(r)$ within sub-samples selected to be intrinsically small $\lambda_{-}$, or intrinsically large $\lambda_{+}$, in order to assess the fractional difference in clustering $\dgg(r)$ across the FP.

We found that elliptical galaxies with intrinsically small effective radii $\lambda_{-}$ are significantly more clustered ($5.6\sigma$) over scales $\sim0.5-17\mpch$ than their large counterparts in the simulation, in agreement with the findings of \cite{joachimi} and \cite{Singh2020}. However, for our fiducial planes, the significance of this signal is diminished ($0.7\sigma$) upon removal of the lower-mass elliptical galaxies with $10^{10}<M_{*}/\msol<10^{10.5}$.

We explored the dependence of FP residuals, and other elliptical galaxy properties, on the isotropic density of their local environments, estimated according to the galaxy number density contrast $\dgr$ in spheres of radius $S$. We saw that the conservative stellar mass threshold removes compact, fainter, lower-dispersion ellipticals from high density environments, thereby flattening the density gradient between intrinsically small $\lambda_{-}$ and large $\lambda_{+}$ samples, and rendering $\dgg$ insignificant.

We detected significantly positive $\dgg$ ($5.9\sigma$), $\xigl$ ($5.2\sigma$), and $\xill$ ($9\sigma$) signals for the Full spiral FP, showing that, conditional on the presence of lower-mass spirals, intrinsically large spirals are significantly more clustered, and exist in regions of higher density, than small spirals. However, $\dgg$  and $\xigl$ weaken to insignificance upon removal of the lower-mass $M_{*}<10^{10.5}\msol$ subset ($0.6\sigma$ and $1.6\sigma$, respectively), whilst the Masscut $\xill$ remains statistically significant, at $3.4\sigma$. This suggests that intrinsic fluctuations in spiral galaxy virial radii are consistently spatially correlated in the simulation, as reckoned by our fiducial spiral FP.

Analysis of the environmental dependence of spiral galaxies' intrinsic sizes and other properties revealed that the mass-selection, affecting $\sim75\%$ of the sample, preferentially removes absolutely and intrinsically small-radius objects from low-density environments. The consequence for $\dgg$ and $\xigl$ is that the signal-to-noise is greatly diminished, whilst the auto-correlation $\xill$ relies less upon a broad range of environments for its significance.

We proposed a tentative interpretation of our measured correlations and intrinsic size-isotropic density relations in the context of the findings of \cite{Welker2017}. We suggested that compact ellipticals in the simulation form by wet mergers of spiral galaxies, which trigger gas compaction and destruction of the disc structure whilst galaxies migrate from filaments into nodes. 
Conversely, we speculated that large spirals are the result of elliptical objects grown large through mergers on their approach to filament centres and nodes, which then regress to disc-like morphologies under cold gas flows and minor mergers. 
We recommend that future work undertake a detailed consideration of the anisotropic cosmic web environment as it influences accretion mechanisms, and thus the galaxy size-mass-morphology distribution, across cosmic time.

We also conducted a mock estimation of the ISC contamination $\Sigma_{\lambda}(r)$ affecting a convergence-based measurement of the projected surface mass density $\Sigma(r)$, as traced by FP residuals \citep{Huff2011}. Assuming our $z\sim0$ ISCs to be representative out to $z\sim0.2$, we estimated the contribution to $\Sigma(r)$ that could arise due to over-estimation of source redshifts at the lens plane, such that `background' galaxies exhibit size fluctuations that are correlated with the foreground structure, in the absence of any gravitational lensing.

We found that the contamination is likely to be of comparable amplitude to the cosmological signal for large separations $r\gtrsim10\mpc$, but subdominant on smaller scales. The spiral- and elliptical-derived contaminations are of similar form but opposite sign, offering a promising route to cross-check the modelling of galaxy type-dependent ISCs under a colour-split magnification analysis \citep[see][for a colour-split consistency analysis in the context of cosmic shear]{Li2021}.

We constructed a simple phenomenological model for the type-specific intrinsic size fields, assuming them to be linearly proportional to the density field \citep{joachimi,Alsing2015} and thus each described by single parameters $B_{\rm{spi}},B_{\rm{ell}}$, such that $\xill=B^{2}\xi_{\delta\delta}$. Fitting to our measured $\xill$ correlations, we found a strong tendency for a non-zero spiral intrinsic size-density coupling $|B_{\rm{spi}}|$ as reckoned by the Full FP sample, weakening to a $\sim2\sigma$ non-zero preference after the mass-selection, but having a consistent amplitude. For the fit to the Full elliptical FP sample $\xill$, we found $|B_{\rm{ell}}|>0$ at almost $3\sigma$, whilst the Masscut constraint is fully consistent with a zero amplitude, having run into the prior.

We converted these models into expectations for intrinsic-intrinsic $C_{\lala}(\ell)$, and intrinsic-convergence $|C_{\lk}(\ell)|$ angular power spectra that could contaminate the cosmological convergence signal $C_{\kk}(\ell)$, assuming that our low-redshift ISCs can be extrapolated to $z\sim0.4,0.8$. For the Masscut FP case, we saw that the resulting systematic $\lala,\lk$ contributions from spirals are likely to dominate over the convergence signal at $68\%$ confidence. This increases to a highly significant dominance for the Full FP case, where $\lk$ $(\lala)$ are factors of $\sim2$ ($\sim$several) larger than $\kk$ with small uncertainties.

Meanwhile, we saw that elliptical $\lk$ contributions are consistent with those predicted by \cite{joachimi} for the Masscut case, and thus comparable to the convergence signal. The corresponding $\lala$ contribution is somewhat weaker, perhaps comparable with $\kk$ at the upper-end of the 68\% confidence interval -- however, the 95\% lower-bound is comfortably consistent with a null signal for all Masscut elliptical predictions. The Full FP-derived constraint upon the linear amplitude of elliptical ISCs exceeds that found by \cite{joachimi} at 68\% confidence, but is consistent at 95\%. Thus the elliptical $\lala,\kl$ contributions predicted from the Full FP are at least comparable to the convergence $\kk$, and likely to be dominant at 68\% confidence, having amplitudes $2-3$ times larger than the cosmological signal. Our results therefore suggest that the impact of intrinsic size correlations upon lensing magnification statistics is likely to require modelling and marginalisation, in order to avoid biases in cosmological parameter inference.

We note that our FP residuals showed variably significant correlations with FP variables and other galaxy properties, including measured radii, stellar mass, and surface area, with spirals showing the strongest $\lambda$-property correlations. We investigated these tendencies in the appendix, considering extended FPs with additional variables: the projected surface area for spirals, and the stellar mass for ellipticals. These extended FPs exhibited rms scatters $1.4-1.7$ times smaller than the fiducial versions, and were able to soften or negate many of the $\lambda$-property correlations.

ISCs measured for the extended FP residuals were seen to be qualitatively consistent with the those from the fiducial FPs, with the detection significance rising to $>3\sigma$ for the Masscut elliptical \& spiral $\dgg$, and Full elliptical $\xigl,\xill$ signals. The exception was the Masscut spiral $\xill$ signal, which was only detected at $2\sigma$ according to the extended FP. Thus the observed spiral intrinsic size auto-correlation may have been contaminated by $\lambda$-property correlations within the Masscut FP sample, and our corresponding phenomenological model predictions for the systematic contributions $C_{\lala}(\ell),|C_{\lk}(\ell)|$ should be considered as upper limits.

A future analysis of intrinsic size correlations in hydrodynamical simulations can improve upon this work first by considering more observationally motivated planes, e.g. using the disc scale-length for spiral FPs \citep{Shen2002}, and by carefully optimising the planes for the avoidance of FP residual-galaxy property correlations that might create spurious signals or complicate interpretations.

An extension to higher-redshift snapshots could assess the evolution of the ISC effect, and test the validity of the strong assumptions and extrapolations made here. Additionally making use of precise knowledge of the underlying, anisotropic dark structure of the cosmic web, combined with galaxy merger trees and larger simulated volumes \citep[e.g. Horizon Run 5;][]{Lee2021c}, the primary drivers of ISCs could be identified and incorporated into modelling.

More physically motivated prescriptions for ISCs include a unified linear model for spirals and ellipticals \citep{Ghosh2020}, and an effective field theory model \citep{Vlah2019}, and the calibration of these would be greatly aided by the identification of a unified Fundamental Plane describing both morphological galaxy types \citep{Ferrero2021}. Such improvements to the modelling of intrinsic galaxy size correlations could shed new light upon the formation and evolution of galaxies in distinct environments within the cosmic web, as well as facilitating direct forecasting of cosmological parameter biases due to ISC contamination of lensing magnification statistics.

\vspace{0.3cm}
{\small {Acknowledgements: the authors acknowledge useful discussions with Andrej Dvornik, Robert Reischke, Corentin Cadiou, and Christophe Saulder. This work is part of the Delta ITP consortium, a program of the Netherlands Organisation for Scientific Research (NWO) that is funded by the Dutch Ministry of Education, Culture and Science (OCW). This work has made use of the HPC resources of CINES (Jade and Occigen supercomputer) under the time allocations 2013047012, 2014047012 and 2015047012 made by GENCI. This work is partially supported by the Spin(e) grants ANR-13-BS05-0005 (\url{http://cosmicorigin.org}) of the French Agence Nationale de la Recherche and by the ILP LABEX (under reference ANR-10-LABX-63 and ANR11-IDEX-0004-02). Part of the analysis of the simulation was performed on the DiRAC facility jointly funded by STFC, BIS and the University of Oxford. We thank S. Rouberol for running smoothly the Horizon cluster for us.
}}

\section*{Data Availability}

Simulation products can be shared upon reasonable request via the collaboration's webpage: \url{https://www.horizon-simulation.org/data.html} or by getting in touch with the corresponding author.

\bibliographystyle{mnras}
\bibliography{IntrinsicSizes}

\appendix

%--------------------------------------------------------------------

\begin{appendix} %First appendix

\section{Extended Fundamental Planes}
\label{sec:extended_fundamental_planes}

\begin{table*}
    \centering
    \begin{tabular}{lccccccc}
\hline
\hline
FP/lens sample & $N$ & $\langle L \rangle/L_{\rm{piv}}$ & $a\,|\,\alpha$ & $b\,|\,\beta$ & $d\,|\,\delta$ & $c\,|\,\gamma$ & $\sigma_{\rm{FP}}$ \\
\hline
Elliptical FP & 6254 & 0.55 &                         $-0.146\pm0.309$ & $-0.348\pm0.055$ & $0.421\pm0.092$ & $0.737\pm0.013$ & $0.0453$ \\
\hline
Elliptical FP ($>10^{10.5}M_{\odot}$) & 3684 & 0.84 & $-0.184\pm0.424$ & $-0.354\pm0.072$ & $0.436\pm0.142$ & $0.838\pm0.017$ & $0.0495$ \\
\hline
Spiral FP & 26215 & 0.24 &                        $-0.145\pm0.017$ & $-0.485\pm0.105$ & $0.331\pm0.029$ & $-1.497\pm0.006$ & $0.0507$ \\
\hline
Spiral FP ($>10^{10.5}M_{\odot}$) & 6394 & 0.59 & $-0.182\pm0.035$ & $-0.891\pm0.229$ & $0.291\pm0.054$ & $-1.294\pm0.013$ & $0.0557$ \\
\hline
\hline
    \end{tabular}
    \caption{Sample details and extended Fundamental Plane constraints for elliptical and spiral galaxy samples defined in the Horizon-AGN simulation. Columns are the same as in Table \ref{tab:sample_details}, but now including a fourth variable $d|\delta$ for the extended Fundamental Planes described in Appendix \ref{sec:extended_fundamental_planes}, and without the Lens samples, which are not fitted with Fundamental Planes.}
    \label{tab:sample_details_extendedFPs}
\end{table*}

\begin{figure}
    \centering
    \includegraphics[width=\columnwidth]{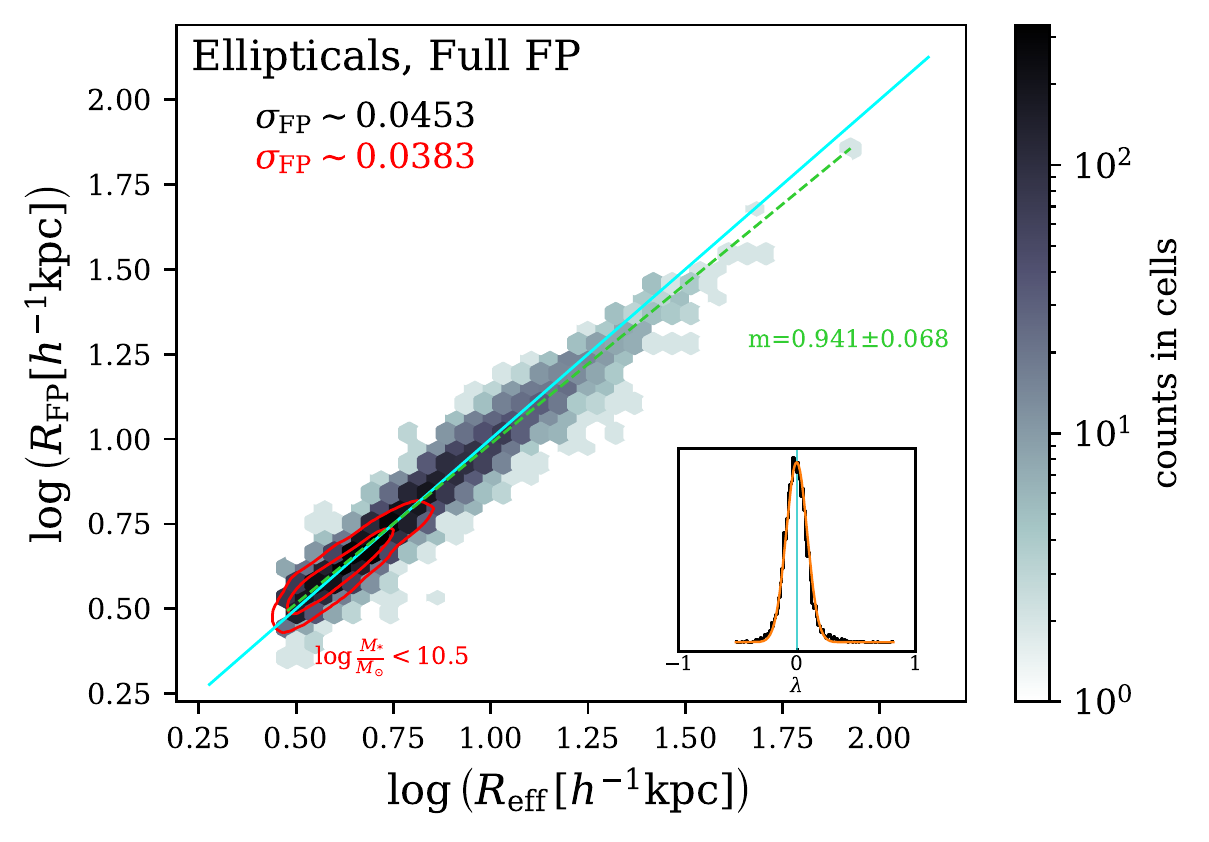}
    \caption{The Full elliptical sample extended Fundamental Plane (analogous to top-left panel of Fig. \ref{fig:fittedFPs}), given by Eq. \ref{eq:extended_elliptical_FP}. The $1:1$ relation is shown in cyan, a linear fit to all points on the plane, and its gradient $m$, are shown in green (Sect. \ref{sec:fundamental_planes_results}), and contours illustrating 68\% and 95\% of $M_{*}<10^{10.5}\msol$ objects (Sect. \ref{sec:simulation}) are shown in red. The addition of stellar mass as a fourth Fundamental Plane variable is seen to effectively reduce the scatter and tilting of the plane.}
    \label{fig:extended_FPs}
\end{figure}

\begin{figure*}
    \centering
    \includegraphics[width=\textwidth]{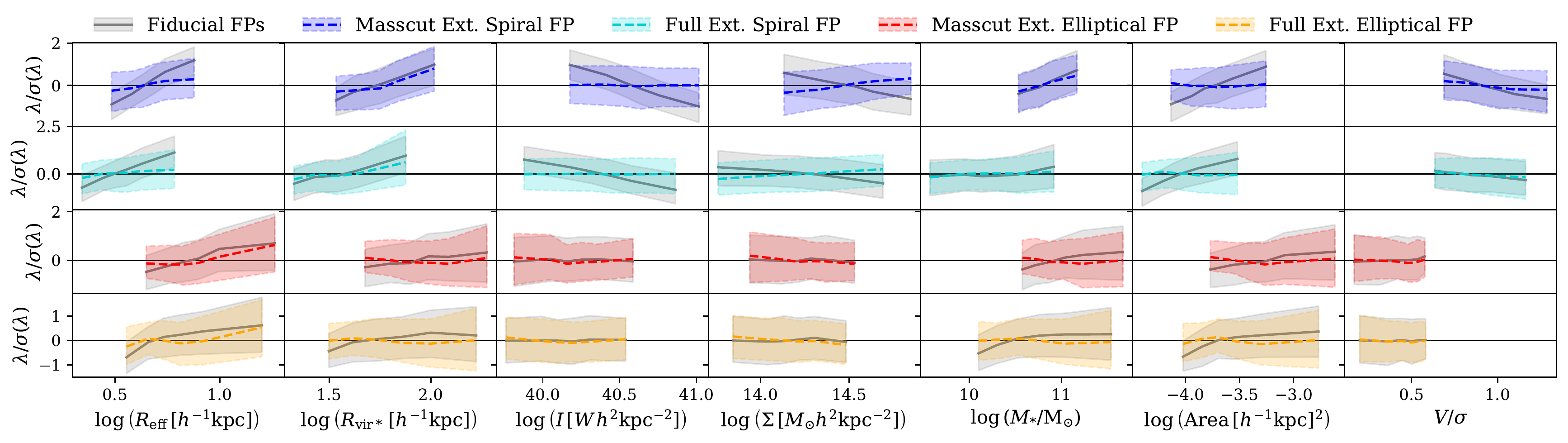}
    \caption{Intrinsic sizes $\lambda$, as estimated from Fundamental Plane residuals (Sect. \ref{sec:fundamental_planes}), normalised by their respective standard deviations $\sigma(\lambda)$ (to ensure a clear comparison between planes with different scatters), vs. a selection of galaxy properties from Fig. \ref{fig:lambda_vs_gxy_props} which displayed non-zero correlations for the fiducial planes. Those relations are reproduced here in grey, whilst colours give the equivalent relations for the extended planes. Rows give, respectively, the Masscut spiral sample, the Full spiral sample, the Masscut elliptical sample, and the Full elliptical sample. In each panel, the extended plane is seen to reduce the degree of correlation between plane residuals $\lambda$ and the respective galaxy property, or else leave it unchanged.
    }
    \label{fig:extended_FPs_lambda_gxy_props}
\end{figure*}

\begin{figure}
    \centering
    \includegraphics[width=\columnwidth]{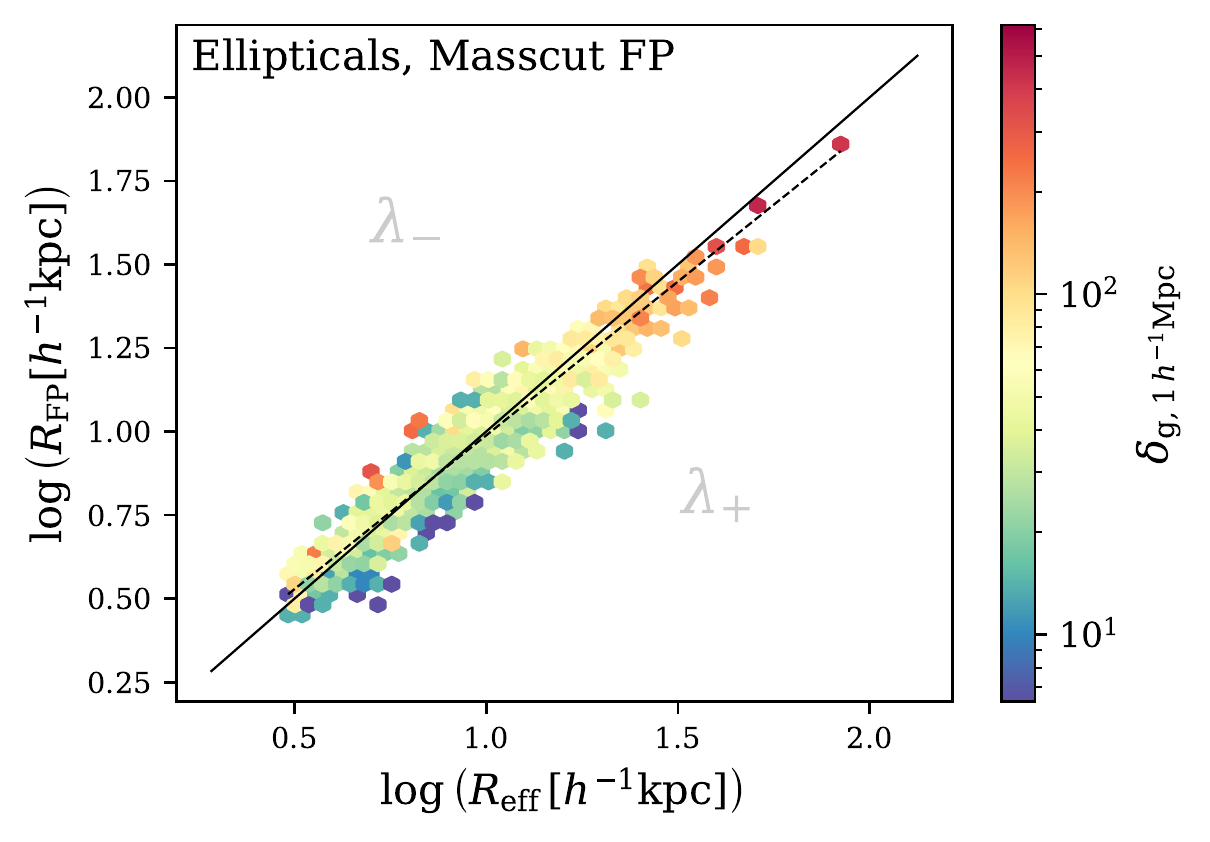}
    \caption{The Masscut elliptical sample extended Fundamental Plane, coloured according to the local galaxy density contrast $\dgr$ estimated in spheres of radius $S=1\mpch$ (analogous to top-left panel of Fig. \ref{fig:FPs_by_density}). The $1:1$ relation, the linear fit to all points on the plane (Sect. \ref{sec:fundamental_planes_results}), and the contours illustrating the distribution of $M_{*}<10^{10.5}\msol$ objects (Sect. \ref{sec:simulation}), are reproduced from Fig. \ref{fig:fittedFPs}, here in black. Annotations $\lambda_-$ and $\lambda_+$ denote the sides of the plane corresponding to intrinsically small and large objects, respectively. The local density gradient from intrinsically small $\lambda_-$ to large $\lambda_+$ objects is stronger for the extended than for the fiducial FP (Fig. \ref{fig:FPs_by_density}; bottom-left panel), and results in a significantly  negative detection of $\dgg$ (Eq. \ref{eq:delta_gg} ); that is, intrinsically small ellipticals defined on the extended Masscut FP are significantly more clustered than their large counterparts, which was not seen for the fiducial Masscut plane.
    }
    \label{fig:extended_FPs_by_density}
\end{figure}

\begin{figure}
    %\centering
    \includegraphics[width=0.9\columnwidth]{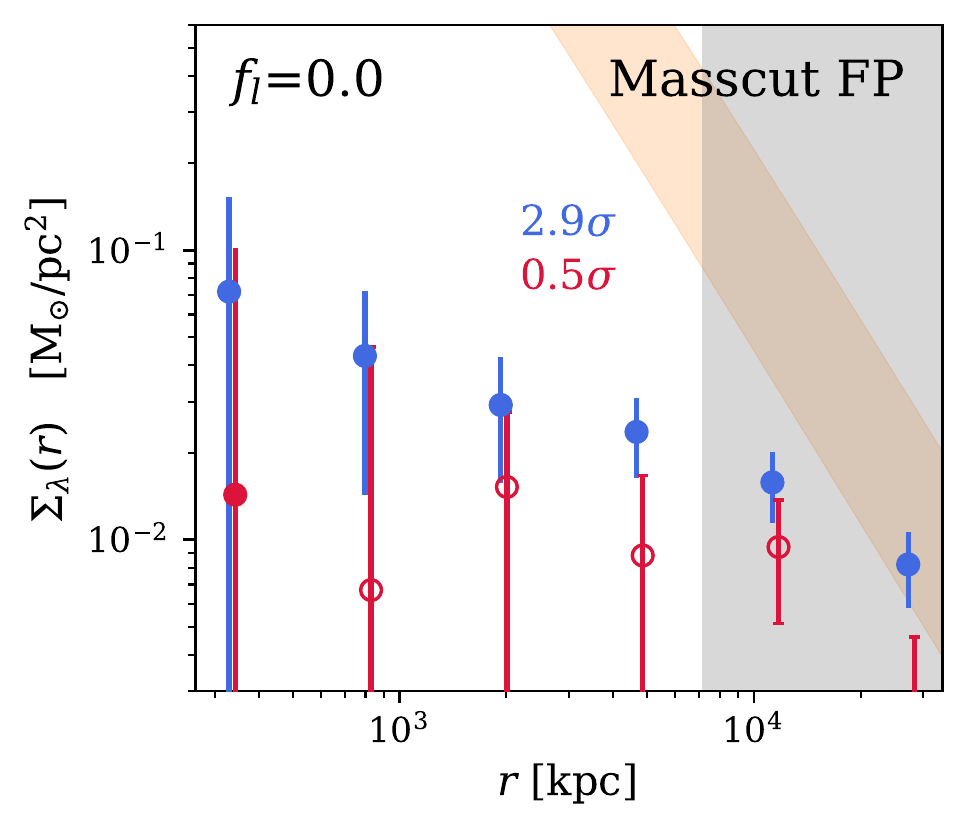}
    \caption{Predicted intrinsic contributions $\Sigma_\lambda(r)$ to the projected surface density $\Sigma(r)$, as measured from angular galaxy position-size correlations in the presence of catastrophic photometric redshift outliers, as in Fig. \ref{fig:Sigma_lambda} (see Sect. \ref{sec:density_magnification_contamination}), but now computed with residuals $\lambda$ from the Masscut extended Fundamental Plane, and showing only the interloper fraction $f_l=0$ panel. The inclusion of the spiral surface area as a fourth Fundamental Plane variable results in a greatly increased significance for the spiral $\Sigma_\lambda(r)$ contribution.
    }
    \label{fig:Sigma_extendedFPs}
\end{figure}

We outline here the extended Fundamental Planes described in Sect. \ref{sec:ISCs_summarised}, which are intended to reduce the correlations between FP residuals $\lambda$ and galaxy properties, shown for the fiducial planes in Fig. \ref{fig:lambda_vs_gxy_props}. To that end, we include a fourth variable in each FP: the stellar mass $M_{*}$ for ellipticals, and the projected surface area $A$ for spirals. The extended planes are thus given as
\begin{equation}
    \log\reff = a\,\log\sigma + b\,\log{}I + d\,\log{}M_{*} + c \quad,
    \label{eq:extended_elliptical_FP}
\end{equation}
for ellipticals, and
\begin{equation}
    \log\rvir = \alpha\,M_r + \beta\,\log{}V_{\rm{c}} + \delta\,\log{}A + \gamma \,,
    \label{eq:extended_spiral_FP}
\end{equation}
for spirals. The resulting coefficients are given in Table \ref{tab:sample_details_extendedFPs}, where one sees that the velocity dispersion slope is heavily flattened and reversed in sign, also picking up a much larger scatter such that $a$ is consistent with zero, whilst the new stellar mass slope $d$ is positive. For spirals, absolute magnitude and circular velocity slopes $\alpha,\beta$ are softened, and the new surface area slope is positive.

The agreement between Masscut and Full FP coefficients is improved for $a,b,\alpha,$ and $\beta$, with $\alpha$ becoming consistent at $1\sigma$, and $\beta$ at $2\sigma$, whilst intercepts $c,\gamma$ are almost unchanged. The additional coefficients $d,\delta$ are also seen to be consistent at $1\sigma$ between the Masscut and Full planes. The rms scatters of extended FP residuals $\sigma_{\rm FP}$ are reduced by factors of $\sim1.5$ for ellipticals, and $\sim1.6$ for spirals, compared with the fiducial FPs (Table \ref{tab:sample_details}). Each extended plane also shows a significantly reduced tilt with respect to its fiducial counterpart, which can be clearly seen for the Full elliptical FP in Fig. \ref{fig:extended_FPs}. The gradients of linear fits to each extended plane (green dashed line(s) in Fig. \ref{fig:extended_FPs}, or in Fig. \ref{fig:fittedFPs} for the fiducial planes) are consistent with unity at $<1\sigma$ for ellipticals, $<2\sigma$ for Masscut spirals, and $<3\sigma$ for the Full spiral sample.

A selection of FP residual property correlations are shown in Fig. \ref{fig:extended_FPs_lambda_gxy_props}, where grey curves give the fiducial relations, reproduced from Fig. \ref{fig:lambda_vs_gxy_props}, and coloured curves give those for the extended FPs -- each is renormalised here by the standard deviation $\sigma(\lambda)$, to aid with the comparison. One sees that the extended FPs either reduce the $\lambda$-property correlations or leave them unchanged. Worrisome persistent correlations are those with $\rvir,M_{*}$ for spirals, and $\reff$ for ellipticals -- future work should explore further methods for the mitigation of these correlations, such that derived intrinsic size correlations will be as free from contamination as possible.

We measure ISCs for the extended FPs as described in Sect. \ref{sec:intrinsic_size_correlations}, and find signals broadly consistent with those measured for the fiducial FPs. Extended FP signals, however, exhibit generally greater significance of detection, with the both the Masscut and Full elliptical $\dgg$ signals becoming highly significant at $>8\sigma$, having similar form and amplitude to the fiducial Full FP signal in Fig. \ref{fig:ISCs}. The Full elliptical FP $\xigl$ also becomes statistically significant ($3\sigma$) whilst maintaining its form and amplitude, as does the Full elliptical $\xill$ signal ($3.3\sigma$) though with a slight reduction in amplitude, particularly at small scales. The Masscut elliptical $\xigl,\xill$ signals also maintain their form whilst rising in significance, but remain at $<2\sigma$.

For spirals, the extended Masscut FP $\dgg$ is detected at $3.5\sigma$, at a slightly higher (but consistent) amplitude than the fiducial signal. Other spiral signals are very similar in form and significance between the fiducial and extended FPs, with the exception of $\xill$. For the extended Masscut and Full FPs, the significance of detection of $\xill$ drops by $3.4\sigma\rightarrow2\sigma$, and $9\sigma\rightarrow5.2\sigma$, respectively, accompanied by reductions in amplitude by factors of $\sim3$, whilst the shapes of the signals are maintained. It may then be that the fiducial $\xill$ signals are partially contaminated by correlations between FP residuals $\lambda$ and galaxy properties; for example $\reff$, or the surface area and derived quantities $I,\Sigma$, each of which is largely erased by the extended FP (Fig. \ref{fig:extended_FPs_lambda_gxy_props}). We leave a detailed investigation of these trends to future studies of spiral Fundamental Planes. 

For now, we remark that the net effect of the extended FPs upon $\xill$ signals is to lower their amplitudes (except for Masscut ellipticals, which are in either case consistent with a null detection). This causes agreement to improve between our phenomenological model fits to elliptical $\xill$ (Sect. \ref{sec:magnification_magnification_contamination}), and those of \cite{joachimi} to the projection of $\xigl$ in observational data, for both Masscut and Full FPs, where the latter becomes consistent at 68\% confidence. The $|B_{\rm spi}|$ constraint from the Masscut $\xill$ signal, meanwhile, becomes consistent with zero; the predictions for contamination of $C_{\kk}$ in the upper panels of Fig. \ref{fig:kappakappa_masscut} should thus be taken with moderation, and considered as somewhat pessimistic. For the Full FPs, the picture is largely unchanged, but for the reductions in amplitudes. The extended Full FPs therefore predict elliptical, rather than spiral, ISCs to be the dominant systematic contribution to $C_{\kk}$, but with a larger uncertainty than the spiral contribution. Each of the extended FP elliptical, spiral, $\lala$, and $\lk$ contributions are within a factor $\sim2$ of the cosmological signal.

The increased significance of the Masscut $\dgg$ signal for elliptical galaxies can be seen in Fig \ref{fig:extended_FPs_by_density} to derive from the clearer segregation of intrinsically large ellipticals into regions of lower density (as characterised within spheres of radius $S=1\mpch$), compared to that seen in Fig. \ref{fig:FPs_by_density}. Similar trends are seen on large isotropic smoothing scales for spiral galaxies, consistent with our suggestions that intrinsic size gradients are apparent across different environments (Sect. \ref{sec:ISCs_summarised}) for ellipticals (nodes vs. filaments) and spirals (filament centres vs. outskirts and the field), and the reduced tilting of both planes also aids with clearer intrinsic size segregation.

Lastly, we note that the extended FPs yield small differences in our mock measurements of the ISC contamination $\Sigma_\lambda(r)$ of the projected surface density $\Sigma(r)$, as measured by FP residuals \citep{Huff2011}. Masscut signals are seen to be of generally higher significance, and whilst the elliptical contribution remains at $\leq0.5\sigma$, the spiral contribution rises to almost $3\sigma$ when the interloper fraction $f_l$ (Sect. \ref{sec:density_magnification_contamination}) is catastrophically underestimated, as can be seen in Fig. \ref{fig:Sigma_extendedFPs}. For the Full FPs, significances are more similar to the fiducial case when $f_l$ is seriously underestimated, but drop faster as the true value $f_l=1$ is approached, at which point both contributions drop to $<1\sigma$ significance. Our conclusions are thus maintained in the sense that an underestimation of $f_l$ could lead to significant contamination of $\Sigma(r)$ at large scales, whilst we might be more optimistic about the possibility of correcting for the intrinsic contributions with accurate characterisations of $f_l$. More exploration of these potential contaminants with optimised Fundamental Planes could be of great value.

\end{appendix}

%%%%%%%%%%%%%%%%%%%%%%%%%%%%%%%%%%%%%%%%%%%%%%%%%%

% Don't change these lines
\bsp	% typesetting comment
\label{lastpage}
\end{document}